\documentclass[prc,preprint,showpacs,tightenlines,showkeys,nofootinbib]{revtex4}
\usepackage{graphicx}

\newcommand{\beq}{\begin{equation}}
\newcommand{\eeq}{\end{equation}}
\newcommand{\bea}{\begin{eqnarray}}
\newcommand{\eea}{\end{eqnarray}}
\newcommand{\ben}{\begin{eqnarray*}}
\newcommand{\een}{\end{eqnarray*}}

\newcommand{\simlt}{\stackrel{<}{{}_\sim}}
\newcommand{\simgt}{\stackrel{>}{{}_\sim}}

\newcommand{\me}{\mathrm{e}}

\def\vec#1{{\bf #1}}

\begin{document}

\title{Precise numerical results for limit cycles\\ in the quantum 
        three-body problem}

\author{R.~F.\ Mohr\thanks{{\tt rfmohr@osc.edu}},
        R.~J.~Furnstahl\thanks{{\tt furnstahl.1@osu.edu}},                
        R.~J.~Perry\thanks{{\tt perry.6@osu.edu}},      
        and K.~G.~Wilson\thanks{{\tt kgw@pacific.mps.ohio-state.edu}}}     
\affiliation{Department of Physics,\\ The Ohio State University, 
  Columbus, OH\ 43210}

\author{H.-W. Hammer\thanks{{\tt hammer@phys.washington.edu}}}
\affiliation{Helmholtz Institut f{\"u}r Strahlen-und Kernphysik (Theorie),\\ 
    Universit{\"a}t Bonn, 53115 Bonn, Germany\\}

\date{September 26, 2005}

\begin{abstract}
The study of the three-body problem with short-range attractive
two-body forces has a rich history going back to the 1930's. Recent
applications of effective field theory methods to atomic and nuclear
physics have produced a much improved understanding of this problem,
and we elucidate some of the issues using renormalization group ideas
applied to precise nonperturbative calculations. These calculations
provide 11-12 digits of precision for the binding energies in the
infinite cutoff limit. The method starts with this limit as an
approximation to an effective theory and allows cutoff dependence to be
systematically computed as an expansion in powers of inverse cutoffs
and logarithms of the cutoff. Renormalization of three-body bound
states requires a short range three-body interaction, with a coupling
that is governed by a precisely mapped limit cycle of the
renormalization group. Additional three-body irrelevant
interactions must be determined to control subleading dependence on the
cutoff and this control is essential for an effective field theory
since the continuum limit is not likely to match physical systems ({\it
e.g.}, few-nucleon bound and scattering states at low energy). Leading
order calculations precise to 11-12 digits allow clear identification
of subleading corrections, but these corrections have not been
computed.
\end{abstract}
\bigskip

\newpage
\maketitle


\section{Introduction}
\label{intro}

It has been known since the 1930's that a non-relativistic three-body
system with short-range two-body potentials has a peculiar behavior,
with a binding energy that is unexpectedly large. In fact the binding
energy becomes infinite if the range of the attractive two-body
potential is shrunk to zero \cite{Tho35}. For over 30 years it has also
been known that the three-body bound state spectrum exhibits an Efimov
effect \cite{Efi70,Efi71}. There are an increasing number of three-body
bound states as the two-body effective range is reduced or as a high
momentum cutoff is increased,  as long as the two-body bound state
energy is held fixed. For sufficiently large cutoff, $\Lambda$, the
ratio of the energies of two successive three-body bound states
approaches an analytically fixed limit of approximately 515.035
for energies much deeper than the two-body binding energy
\cite{Albe81}.  

        More recently, Bedaque, Hammer, and van Kolck have shown that
the short-range three-body problem is renormalizable, but only if a
point-like three-body interaction is added \cite{BHK99,BHK99b,BHK00}.
The dimensionless version of the three-body coupling strength [which
they denote by $H(\Lambda)$ but we denote by $G_3(\Lambda)]$ has an
unexpected dependence on  $\Lambda$, namely $G_3(\Lambda)$ has a
periodic dependence on $\ln\Lambda$. In each period the value of
$G_3(\Lambda)$ steadily increases until it reaches plus infinity. It
jumps to minus infinity and then steadily increases again. This
peculiar behavior provides a rare example of a renormalization group
limit cycle \cite{Wil00, Wil71, Glazek02}, with the exciting possibility that
scaling behavior near a limit cycle might be studied experimentally
\cite{Bra04}.

Unfortunately, the two experimental candidates to be Efimov-like,
namely the triton and the atomic helium trimer, do not exhibit the
infinite set of shallow Efimov states because that would require the
nucleon-nucleon S-waves or helium dimer systems to have a bound state
precisely at threshold \cite{Efi70,Efi71}. This is not the case
experimentally. The reality is that the triton has only one bound
state. In the case of the helium trimer numerical calculations for
realistic potential models indicate that the trimer has two bound
states. These departures from the infinite Efimov limit raises the
question: what are the corrections to the Efimov limit in the
three-body system when the two-body system does not have a bound state
precisely at threshold? How large are the corrections in the case that
there are only one or two Efimov states rather than an infinite number?

Ideally, one would answer this question for the most plausible
three-body Hamiltonians that have been proposed for describing the
triton or the helium trimer [see refs. in \cite{Bra04}]. But realistic
three-body Hamiltonians are very demanding to study. Their three-body
eigenfunctions are functions of three variables in the simplest case
(S-states), making both analytical and numerical analysis exceptionally
difficult (although far from impossible). Fortunately, it is possible
to formulate a cutoff form of a two-body zero range potential (a
delta-function potential $V({\bf r})=g \delta^3({\bf r})$) in which the
Schr{\"o}dinger equation wave function can be expressed exactly in
terms of a reduced wave function $\phi(r)$ that depends only on one
radial variable rather than three variables. The Schr{\"o}dinger
equation itself becomes a one-dimensional integral equation for
$\phi(r)$  and the energy eigenvalue $E$ (although with auxiliary
functions that involve integrals over known functions).

The one-dimensional integral equation can be solved numerically and
quickly to machine precision (around 12 decimal places in our
calculations) on a PC, although achieving this high precision involves
careful attention to details of how the reduced wave function and the
integral equation are discretized. High numerical accuracy is
invaluable for testing the validity of analytic but approximate
formulae for the corrections to the Efimov limit.

The integral equation will be formulated and solved for the Fourier
transform $\Phi(p)$ of the reduced wave function. Uniformly valid
expansions ({\it e.g.}, with uniform convergence to the exact solution over 
the entire
range of momenta) are employed to systematically dissect the bound
state equation so that high-momentum behavior is isolated analytically.
Low, intermediate and high momentum scales in the bound-state
wavefunction are isolated by defining approximations to the full
wavefunction that are valid in each of these three regions, and then
showing that these can be assembled to produce an approximation that is
valid for all momenta. The wave-functions valid at low and high momenta
``communicate" through a single phase found in an analytic solution for
the intermediate momenta approximation, all to leading order.

The outcome of the analytic and numerical analysis of this article is
that Efimov three-body bound states have energy eigenvalues $B_3^{(n)}$
whose infinite cutoff limits can be precisely computed, and a simple
renormalization group analysis suggests the leading corrections are of
order $\sqrt{B_2} ~{\rm ln}(\Lambda)/\Lambda$ and $B_3^{(n)}~ {\rm
ln}(\Lambda)/\Lambda^2$ relative to $B_3^{(n)}$ itself. $B_2$ is the
binding energy of the two-body state near threshold. But these leading
corrections have coefficients which depend on the details of the two-
and three-body interactions in the Hamiltonian, such as the detailed
shape of the two-body cut off potential, which suggests that the actual
size of corrections to the Efimov limit for the triton or helium trimer
could depend on the details of the Hamiltonians that best characterize
these systems.

There is a second reason for writing this article. The Hamiltonian
studied in this article exhibits renormalization group limit cycle
behavior, at least to the numerical accuracy achieved. We develop a
renormalization group description of the Hamiltonian that takes into
account the leading corrections to the Efimov limit as derived from the
integral equation. The renormalization group description takes the form
of an extended Gell-Mann-Low analysis \cite{Gell54} involving two-body
and three-body coupling parameters. The extended Gell-Mann-Low analysis
provided here accounts for the limit cycle and for the leading
corrections that are of relative order $\sqrt{B_2}~{\rm
ln}(\Lambda)/\Lambda$ as well as corrections of order
$\sqrt{B_2}/\Lambda$. This requires two two-body couplings and one
three-body coupling. The analysis reported here is somewhat similar to
the analyses produced in effective field theory (EFT). However the
analysis offered here differs from EFT in both the two- and three-body
sectors because it starts from a non-perturbative fixed point in the
two-body sector and from the three-body limit cycle in the three-body
sector. There are no new outcomes in our approach that differ from the
EFT analysis for the two-body sector. It is only in the three-body
sector that, due to the limit cycle, the renormalization group analysis
provides a more straightforward approach than anything published to
date using EFT.

A major concept of the renormalization group is the concept of
universality. The concept of universality was developed
for the case of renormalization group fixed points and has two parts.
The first part (due mostly to Franz Wegner \cite{Weg72}) is an analysis
of infinitesimal departures from the fixed point Hamiltonian, such as
infinitesimal departures from the fixed point of the two-body
Hamiltonian. Franz Wegner also formulated a classification of operators
that are relevant, marginal or irrelevant. This classification requires 
adjustments before it applies to
infinitesimal departures from a renormalization group limit cycle. One
adjustment is that whenever there is a limit cycle, there is an
infinitesimal marginal operator whose role is to correspond to an
infinitesimal shift on the limit cycle itself. The coefficient of this
``limit cycle shift" operator is an angular variable $\theta$, and all
physics is invariant to changes in $\theta$ by $2 \pi$. Other
adjustments are more minor.

We will present our analysis in a number of stages, with analysis of
the integral equation postponed until the last two stages. The stages
are as follows:

\begin{itemize}

\item The two-body renormalization group, neglecting irrelevant
operators (Section~II).

\item The two-body renormalization group with the leading irrelevant
operator (Section~III).

\item The three-body renormalization group, neglecting irrelevant
operators, and the possibility of a limit cycle (Section~IV).

\item The three-body renormalization group with the leading irrelevant 
operator (Section~V).

\item Derivation of the integral equation in the presence of a
Gaussian-like cutoff (Section~VI).

\item Methods of solution of the integral equation: method for analysis
of limiting behavior for large $\Lambda$, applied to a simplified
example (Section~VII).

\item A renormalized equation in the three-body case (Section~VIII).

\item Discretization of the integral equations with exponentially small
errors (Sections~IX and X).

\item Analytic and numerical results (Section~XI).

\end{itemize}

The initial discussions of the three-body renormalization group
equations draw on results reported later, so a full understanding of
this work probably requires two passes. Additional details and precise
calculations can be found in Richard Mohr's thesis \cite{Mohr03}. For a detailed
account of previous work on the three-body system with short-range interactions,
see Ref. \cite{Bra04} and references therein.


\section{The Two-Body Renormalization Group}
\label{2bdy_rg}

        We develop the renormalization group equations first for the
two-body sector. These equations serve unchanged as a subset of the
renormalization group equations for the three-body sector. In its
simplest form, excluding irrelevant operators, the two-body
renormalization group equation is for a single dimensionless coupling
constant $G_2(\Lambda)$. The equation for $G_2(\Lambda)$ has a fixed
point solution at a value $G_2^*$ for which the two-body binding energy
is zero. The fixed point solution for $G_2(\Lambda)$ provides a
contrast to the limit cycle solution for $G_3(\Lambda)$, and it is
interesting in its own right because it is a non-free fixed point far
removed from the free fixed point \cite{Bir98}. The existence of this
fixed point leads to scaling behavior that seems unnatural if one seeks
to expand around the free theory \cite{KSW98,KSW98b}. Its existence
brings into question the use of free operators (i.e., powers of fields
and derivatives with dimension given by a free scaling analysis)
instead of operators with good scaling behavior (i.e., eigenoperators
of a linearized RG transformation about the interacting fixed point
\cite{Weg72}.) For a complete RG analysis using a slightly different
formalism, see \cite{Bir98}.

        The bound state equation and the scattering amplitude for a
separable two-body Hamiltonian are known; we report the result as the
starting point for analysis. We consider a two-body Hamiltonian with
two identical Bose particles of mass $M=1/2$ in the center-of-mass
frame. This mass is chosen to simplify the algebra in the three-body
problem. The equivalent one-body Hamiltonian is:
\beq
  H = 2 {\vec p}^2 + V\ ,
\eeq
where we choose the separable two-body potential $V$:
\beq
  V (\vec{p}, \vec{p}^\prime) = -{G_2 \over \Lambda}~ U_\Lambda({\vec
    p})U_\Lambda(\vec{p}^\prime).
    \label{seppot}
\eeq 
$U_\Lambda({\vec p})$ can in principle be any real function of the
three-momentum $\vec p$ and the cutoff $\Lambda$ that goes to zero as
$p$ becomes much larger than $\Lambda$ and to one as $p$ goes to zero, but  
symmetries should be respected if possible. We have written
the interaction using a dimensionless coupling, $G_2$, because we need
to disentangle the scaling dependence of couplings that serve as
coordinates in a space of operators from the scaling dependence of the
operators themselves. It is much simpler to develop RG-improved
perturbation theory about a non-free fixed point using the parameters
that must remain small near the fixed point, and these are the
dimensionless couplings.

The exact bound state equation reduces to
\beq
{\Lambda \over G_2} = \int d \tau {U_\Lambda^2 ({\vec p}) \over 2 {\vec p}^2 
    + B_2}
\ ,
\label{exactbd}
\eeq 
where the integration volume is $d\tau= d^3{\vec p} / (2 \pi)^3$. The
K-matrix is equivalent to the T-matrix but uses standing wave boundary
conditions that produce a principal value prescription for scattering
integrals. It satisfies the integral equation
 \beq
   K(\vec{p}, \vec{p'}; E) = V(\vec{p}, \vec{p'}) 
     + {\cal P} \int \frac{d^3\vec{q}}{(2 \pi)^3}
  {V(\vec{p}, \vec{q}) K(\vec{q}, \vec{p'}) \over E-2 q^2}.
\label{Keq}
\eeq 
For a separable potential this equation is easily solved:
\beq
K(\vec{p}, \vec{p'}; E) = 
   -\frac{U_\Lambda (\vec{p}) U_\Lambda(\vec{p'})}{D(E)}
\ ,
\label{exactK}
\eeq 
where
\beq
    D(E) = { \Lambda  \over G_2} - {\cal P} \int d \tau {U_\Lambda^2 ({\vec p}) 
          \over 2 {\vec p}^2 - E}.
\label{exactD}
\eeq
Once again, this is an exact result, which allows us to identify and
explore the region about the fixed point.

The K-matrix is directly related to the S-wave phase shift and thereby
to the effective range expansion for low energy scattering using
\beq
p ~{\rm cot}(\delta_p) = - { 8 \pi \over K(E)}= -{1 \over a} 
   + {1 \over 2} r_e p^2 + \cdot \cdot \cdot ,
\label{Keffrange}
\eeq
where $K(E)$ is the on-shell K-matrix with $p=p', E=2 p^2$, $a$ is the
scattering length and $r_e$ is the effective range; and we will confine
our attention to S-waves. This relationship is ideally suited to the
development of a Gell-Mann-Low analysis. We take advantage of the fact
that we can solve this two-body problem exactly in the limit $\Lambda
\rightarrow \infty$, using a Gell-Mann-Low analysis to identify all
operators required, rather than attempting an exact Wilsonian RG
analysis \cite{Wil75}.

This limit yields (see for example \cite{jackiw})
\beq
p ~{\rm cot}(\delta_p) = 0 ~,~D(E)=0.
\eeq
The scattering length is infinite, so there is a bound state at zero
energy. A binding energy can only be introduced by adding a scale,
which separates infrared and ultraviolet scaling behavior and which
therefore must be associated with moving away from an ultraviolet fixed
point. We do not provide a full analysis, but for any couplings that
have a continuum limit,
\beq
p ~{\rm cot}(\delta_p) = -\sqrt{B_2/2}.
\label{exacteffrange}
\eeq
This is true for any momentum and any
cutoff because an exact RG produces `physical'
results that are completely independent of the cutoff and which
therefore must be the same as these results in the limit where the
cutoff is taken to infinity. This can be used to understand the limits
of a Gell-Mann-Low analysis. In general, one adjusts the irrelevant
operators in an EFT Hamiltonian so that the effective range expansion
matches data. This requires these couplings to be tuned away from their
continuum RG trajectory, and we know that running the transformation
backward will cause irrelevant deviations from continuum theory to
deviate exponentially from continuum trajectories in the ultraviolet
limit. At some point the irrelevant couplings start to become large and
the RG-improved perturbative analysis breaks down; the theory becomes
unnatural in the parlance of effective field theory. There is also an
issue of how rapidly an expansion of the actual irrelevant operators in
powers of free operators converges, but we do not investigate this
issue here and turn instead to the use of a smooth Gaussian cutoff.

The Hamiltonians we want to study must be approximated as deviations
from the fixed point, and in general we will approximate the fixed
point and the relevant and irrelevant operators using
\beq
U_\Lambda(p) = (1+h_2(\Lambda)~ p^2 + ...) ~{\rm exp}(-p^2/\Lambda^2).
\label{approxU}
\eeq
We can adjust $G_2(\Lambda)$ to fix the binding energy by solving Eq.
(\ref{exactbd}), but we see that it depends on $h_2$. $h_2$ can be
fixed by also insisting that the effective range be zero, which is true
for any continuum theory and introduces the first place at which
effective theories must be allowed to deviate from the continuum limit
for practical application. This requires the introduction of irrelevant
couplings that become of order 1 when $\Lambda=1/r_e$ if the effective
range, $r_e$, is not zero, as we will see below.

It is clearly quite easy to complete a non-perturbative analysis of the
two-body problem, but we will use the Gaussian cutoff and analyze the
problem using the method of uniformly valid expansions, which is a key
tool in our three-body analysis that is presented in Section VII. 
We do not outline the method here but
only point out its appearance, saving the discussion of details. We
will first approximate the fixed point and relevant operator for the
fixed point using
\beq
U_\Lambda(\vec{p}) = \exp (- {\vec p}^2 / \Lambda^2)\ .
\eeq

The equation we seek is an equation for $dG_2/d\Lambda$, expressed as a
function of $G_2$ and $\Lambda$. Since $G_2$ is dimensionless, while
$\Lambda$ is not, the structure of the equation must be
\beq
\Lambda {dG_2 \over d \Lambda} = \beta_{G_2} (G_2)\ ,
\eeq 
where $\beta_{G_2}(G_2)$ is a function to be determined, in this case
from the equation for $1/G_2$. We assume there is a fixed point
$G_2^*$, and the function $\beta_{G_2}(G_2)$ can be determined as an
expansion in powers of $G_2-G_2^*$. The result takes the form:
\beq
\beta_{G_2} (G_2) = - (G_2 - G_2^*) + c_2 (G_2-G_2^*)^2 + \dots\ ,
\eeq 
with
\beq
G_2^* = 4(2 \pi)^{3/2} \quad \mbox { and } \quad
c_2 = {(2 - \pi) \over 2 (2 \pi)^{5/2} }\ .
\eeq 
Note that there is no exact fixed point solution when a Gaussian cutoff
is used, so the value of couplings at the fixed point will change in
our Gell-Mann-Low analysis as we add couplings. We will provide only
the beginning of the analysis that leads to RG equations. The equation
for $1/G_2$ can be rewritten as
\beq
{\Lambda \over G_2} = {1 \over 2 \pi^2} \int^\infty_0 dp \left[ {p^2 \exp (-2
p^2/\Lambda^2)  \over (2p^2 + B_2)}\right]  ,
\eeq 
where we can employ a uniformly valid expansion at next-to-leading
order to write
\beq
{\Lambda \over G_2} \approx {1 \over 2 \pi^2} \int^\infty_0 dp \left[ 
   {p^2 \over 2 p^2+B_2}\left( 1-{2 p^2 \over \Lambda^2} \right)+
   {1 \over 2} {\rm exp}(-2 p^2/\Lambda^2)\left(1-{B_2 \over 2 p^2} \right) - 
   {1 \over 2} \left(1-{2 p^2 \over \Lambda^2}-{B_2 \over 2 p^2} \right) \right].
\eeq 
The first term is the leading order approximation to the integrand in
the region of integration where $p \approx B_2 \ll \Lambda$, the second
term is the approximation in the region where $B_2 \ll p \approx
\Lambda$, and the third term is the approximation in the intermediate
region which cancels the leading errors in the other two approximations
in the regions where they are not valid. This division is not necessary
here, where we know the integrand, but it immediately leads to:
\beq
{1 \over G_2} = {1 \over G_2^*} - {1 \over 8 \sqrt{2} \pi} 
{\sqrt {B_2} \over \Lambda}+{B_2 \over 2 (2 \pi)^{3/2} \Lambda^2}
+{\cal O}\left( \left({\sqrt{B_2}}/{\Lambda}\right)^3\right)\ .
\eeq 
This equation makes it clear that our potential can be used to
approximate both the fixed point and the relevant operator for that
fixed point, where our approximate relevant operator is used to control
the binding energy ({\it i.e.}, the scattering length).


\section{Two-body Renormalization Group with an Irrelevant Operator}
\label{2bdy_rg_irr}

We now expand the renormalization group equations in the two-body
sector to include a coupling coefficient for the leading irrelevant
operator. The purpose of this exercise is to identify and control the
leading cutoff-dependent correction in the effective range expansion.
We introduce a new coupling into the two-body Hamiltonian that, when
varied, enables this correction to be held fixed even as the cutoff
$\Lambda$  is replaced by an effective cutoff $\Lambda^\prime$. 

We want to contrast the form of the leading irrelevant operator in the
two-body sector with the leading irrelevant operator obtained in
perturbation theory. Thus, as a preliminary, we look at irrelevant
operators in first-order perturbation theory for small $G_2$.

        To first order, the K-matrix is
\beq
K(\vec{p}, \vec{p'}) = {G_2(\Lambda) \over \Lambda}~
 U_\Lambda (\vec{p}) U_\Lambda (\vec{p'})\ .
\eeq 
The dominant terms in this expression, for fixed momenta and large $\Lambda$,
are
\beq
K(\vec{p}, \vec{p'}) = {G_2(\Lambda) \over \Lambda} 
          - {G_2(\Lambda) (\vec{p}^2 + \vec{p'}^2) \over \Lambda^3}\ .
\eeq 
In this case, the leading irrelevant terms are the terms involving $\vec{p}^2$
and $\vec{p'}^2$; to keep $K$ constant while changing $\Lambda$, one could
generalize the function $U_\Lambda$ to become
\beq
U_\Lambda(\vec{p}) = \left(1 + h_2 (\Lambda ) {{\vec p}^2 \over \Lambda^2} 
  \right )
\exp (-{\vec p}^2 / \Lambda^2) \ .
\eeq 
Then the renormalization group equations needed to hold $K$ fixed to first
order in an expansion in powers of ${\vec p}^2 / \Lambda^2$ are:
\beq
\Lambda {d G_2 ( \Lambda ) \over d \Lambda} = G_2(\Lambda)\ ,
\eeq 
and
\beq
\Lambda {d \over d \Lambda} \left( G_2( \Lambda ) [ 1 - h_2 (\Lambda )]
\right) = 3 G_2(\Lambda ) [1-h_2 (\Lambda )]\ .
\eeq 

Now we repeat this analysis for the full nonperturbative amplitude. We
identify the leading nonperturbative $\Lambda$-dependent correction to
the scattering amplitude near the fixed point in $G_2$. It is
sufficient to identify the leading correction to the function $D(E)$.
We shall assume that $E$ is of order $B_2$. We can rewrite the equation
for $D(E)$, Eq. (6), using Eq. (\ref{exactbd}) to give:
\beq
D(E) ={\cal P} \int^\infty_0 d\tau \left \{ {U_\Lambda^2(p) \over
(2p^2 + B_2) } - { U_\Lambda^2(p) \over (2p^2 -E) }\right \}\ .
\eeq
This integral is finite as $\Lambda \rightarrow \infty$, so it should
have an expansion in inverse powers of $\Lambda$, although we must allow
for the possibility that logarithms of the cutoff will appear. We can make some
progress before specifying $U_\Lambda$ by using a uniformly valid
expansion for the integrand (as discussed in Section VII)
and the constraint that the cutoff function
goes to zero for large momenta and to one for small momenta. To leading
order the uniformly valid expansion gives us:
\beq
D(E) = {-(E+ B_2) \over 2 \pi^2} {\cal P} \int^\infty_0 dp \left \{ {p^2 \over
(2p^2 + B_2) (2p^2 - E)} +  {U_\Lambda^2(p)\over 4 p^2}  
   - {1 \over 4 p^2} \right \}\ .
\eeq 
Again, the first term is the leading order approximation to the
integrand in the region of integration where $p \approx B_2,E \ll
\Lambda$, the second term is the approximation in the region where
$B_2, E \ll p \approx \Lambda$, and the third term subtracts the
approximation in the intermediate region to cancel the leading errors
in the other two approximations in the regions where they are not
valid. The second and third term together are dominated by values of
$p^2$ of order $\Lambda^2$ because the numerator $1-U_\Lambda^2$ 
vanishes by assumption for smaller values of $p^2$.

The first term is independent of the cutoff and gives $-\sqrt{B_2}/
(8\sqrt{2} \pi)$. The last two terms give $D(E)$ a correction behaving
as $1/\Lambda$, with a coefficient that depends on how $U_\Lambda$
behaves for $p\sim \Lambda$.  For the Gaussian form of $U_\Lambda$ that
we have adopted, the $\Lambda$-dependent term in $D(E)$ is
\beq
{(E+ B_2) \over 2(2 \pi)^{3/2} \Lambda}\ .
\eeq 

If we want to control this cutoff dependence to renormalize the
scattering matrix, we can vary the coefficient of $1/\Lambda$ by
introducing an extra coupling parameter into the definition of
$U_\Lambda$, which thereby would become a new coupling parameter in the
potential $V$. There are many ways to introduce this parameter, all of
which give higher order corrections to $D(E)$ as well as a $1/\Lambda$
correction. We have chosen--quite arbitrarily--to use the form already
specified for $U_\Lambda$ including the extra coupling constant  $h_2
(\Lambda)$. Universality should insure that this arbitrary choice is as
good as any other.

Now one finds that the $\Lambda$-dependent correction to $D(E)$ takes the form
\beq
{[16 - 8h_2 (\Lambda)- h_2^2 (\Lambda)] (E + B_2) \over 
32 (2 \pi)^{3/2} \Lambda} \ .
\eeq 
Holding this term constant yields the renormalization group equation:
\beq
\Lambda {d \over d \Lambda}[16 - 8 h_2 (\Lambda) - h^2_2 (\Lambda )] = 
[16 - 8 h_2 (\Lambda) - h_2^2 (\Lambda)]\ . 
\eeq 
We note that this equation has fixed points at 
\beq
h_2 (\Lambda ) = h^*_2 = 4(-1 \pm \sqrt{2} )\ .
\eeq 
At the fixed points, there is no correction to $D(E)$ in order $1/
\Lambda$. The fact that $h_2^* \ne 0$ indicates that the operator it
multiplies is not really an irrelevant operator. To isolate an
irrelevant operator we must consider deviations of $h_2$ from $h_2^*$.

We close this section with a brief discussion about what happens if one
uses the additional coupling $h_2$ to change the effective range $r_e$
(see Eq. (7)) from its zero continuum value to a finite value in order
to reproduce data and use this Hamiltonian in effective field theory.
As we have shown above, for theories with a continuum limit $r_e=0$;
but to use this as an effective field theory we want to be able to
adjust $r_e$.

We can use Eq.(\ref{exactK}), with $U(p) = (1+h_2~p^2/\Lambda^2)~{\rm
exp}(-p^2/\Lambda^2)$ to find equations for $G_2$ and $h_2$ that allow
us to fix $a$ and $r_e$ at whatever values we please:
\begin{equation}
\frac{1}{G_2} + \frac{1}{8 \pi a} - \frac{\Lambda (3 h_2^2 + 8 h_2 + 16)}
   {128 \sqrt{2} \pi^{3/2}}  =  0 \label{eqn:ar2g},
\end{equation}
\begin{equation}
\frac{1}{4 \sqrt{2 \pi} \Lambda} h_2^2 + \left( \frac{2}{\sqrt{2 \pi} \Lambda} 
   - \frac{2}{a \Lambda^2} \right) h_2 + \left( \frac{2}{a \Lambda^2} 
   + \frac{1}{2} r_e - \frac{4}{\sqrt{2 \pi} \Lambda} \right)  
   =  0 \label{eqn:ar2h}.
\end{equation}
\noindent We can fix $a$ and $r_e$ at cutoff independent experimental
values by using cutoff dependent values for $G_2$ and $h_2$.  In
addition, the error in other $\Lambda$-dependent observable quantities
will now be of ${\cal O}(B_2/\Lambda^2)$. 

Consider what happens to $h_2$ as $\Lambda \rightarrow \infty$. In this
limit the equation for $h_2$ simplifies to
\beq
h_2^2 + 8 h_2 + 2 \sqrt{2 \pi} \Lambda r_e -16 = 0.
\eeq
The solution is
\beq
h_2 = -4 \pm \sqrt{32-2\sqrt{2 \pi} \Lambda r_e}.
\eeq
This result makes it clear that we can use this model as an effective
field theory only for cutoffs that satisfy $\Lambda r_e < 8 \sqrt{2/\pi}$.
Furthermore, we conclude that if we choose $r_e \ne 0$ it is not
possible to completely remove cutoff dependence; because if this were
possible, we would be able to let $\Lambda \rightarrow \infty$. Of
course, this is not a serious limitation as long as $r_e/a \ll 1$; but
if this constraint is not satisfied by the data we wish to model, there
is no reason to believe that we can use this simple effective field
theory.


\section{Limit Cycle for the Three-Body Coupling}
\label{3bdy_rg}

        We now provide the initial analysis showing that a single
coupling constant can approach either a fixed point or a limit cycle as
$\Lambda$  goes to infinity, with no other possibilities, and then
observe that the three-body coupling of Bedaque {\em et al.}
\cite{BHK99,BHK99b,BHK00} exhibits limit cycle behavior. We show that
an infinite sequence of bound states approaching zero energy
geometrically is a natural consequence of a limit cycle.      

        Let $G_3(\Lambda)$ be a three-body coupling constant satisfying
the renormalization group equation
\beq
\Lambda {d G_3 \over d \Lambda} = \beta_{G_3} (G_3)\ .
\eeq 
We assume that $\beta_{G_3} (G_3)$ is differentiable, as it turns out
to be for the example of Bedaque {\em et al.}.  We observe that $G_3
(\Lambda)$ cannot oscillate in value, if the oscillations have either a
finite maximum or a finite minimum, because $\beta_{G_3} (G_3)$ would
have to vanish at either a maximum or a minimum. Such vanishing would
ensure that there are fixed points at the maximum or minimum, and $G_3
(\Lambda)$ would approach one of these fixed points as its limit for
$\Lambda$ going to infinity, rather than oscillate.

There are only two possibilities left for $G_3 (\Lambda)$:

\begin{enumerate}
\item It can change only in one direction, always increasing or always
decreasing, in which case it has a limiting value -- either a finite fixed
point or $\pm \infty$. 

\item It can oscillate but only by jumping from $+ \infty$ to $-\infty$ or vice
versa.

\end{enumerate}

\noindent The formula for $G_3 (\Lambda)$ can be obtained from the 
$H (\Lambda)$ of Bedaque {\em et al.} by multiplying $H (\Lambda)$
with minus one,
\beq
\label{h}
G_3(\Lambda)=      \frac{\sin(s_0\ln({\Lambda}/{\Lambda_\star})-
                   {\rm arctg}(1/s_0))}
                 {\sin(s_0 \ln({\Lambda}/{\Lambda_\star})+
                   {\rm arctg}(1/s_0))}\ ,
\eeq      
where $\Lambda_\star$ is a dimensionful constant that must be chosen
and $s_0 \approx 1.00624$. The additional minus sign is necessary to
translate from the Lagrangian to the Hamiltonian formalism we are
using. $G_3(\Lambda)$ increases while finite, until it reaches $+
\infty$; it jumps to $- \infty$ and then increases again.  We will show
in a later stage that $G_3 (\Lambda)$ must increase rather than
decrease as $\Lambda$ increases, except for the discrete jumps.

Now we use the renormalization group equation to show that in case 2,
$G_3 (\Lambda)$ must exhibit limit cycle behavior. We introduce a new
independent variable:
\beq
t = \ln(\Lambda/\Lambda_0)\ ,
\eeq 
where $\Lambda_0$ is arbitrary, so that the RG equation will be translation
invariant in $t$. The renormalization group equation now reads 
\beq
{dG_3(t) \over dt} = \beta_{G_3} (G_3)\ .
\eeq 
But now by assumption the solution $G_3(t)$ passes from $-\infty$ to $+\infty$
between two values of $t$, say $t_1$ and $t_2$. However, since the
renormalization group equation is translation invariant in $t$, $G_3(t)$  must
repeat itself with a period $t_2 -t_1$, where $t_2 > t_1$. 

        Now we look at the structure of the three-body bound state spectrum. We
consider the case that $G_2$ is $G_2^*$ so that $B_2$ is zero. In this case
there are an infinite set of three-body bound states with energies of the 
form \cite{Efi70,Efi71}:
\beq
B_3^{(n)} = \Lambda^2 F_n [G_3 (\Lambda)]\ .
\eeq 
The factor $\Lambda^2$ provides the energy scale; the functions $F_n$ provide
dimensionless coefficients that depend on the three-body coupling parameter
$G_3(\Lambda)$. The ground state is labeled by $n=0$. When $G_3(\Lambda)$ is
near $-\infty$, $F_0$ is near one (as
will be shown later). Then as $G_3(\Lambda)$ increases towards $+ \infty$,
the $F_n$'s  increase towards zero. {\it As part of keeping low energy physics
unchanged as $\Lambda$ increases, the $B_3^{(n)}$'s for large $n$ do not change.}
However, when $G_3(\Lambda)$ jumps from $+ \infty$ to $- \infty$, a new
ground state emerges and what was $F_n$  now becomes equal to $F_{n+1}$
instead:
\beq
F_n(+ \infty) = F_{n+1} (-\infty)\ .
\eeq 

        Another observation is this. Each time $G_3(\Lambda)$ passes through
zero, the  $F_n$'s must take on the same set of values. Thus suppose that
$G_3(\Lambda_1)$ is zero. Then consider what happens as $\Lambda$ increases
enough so that $G_3(\Lambda)$ increases to $+ \infty$ (with $\Lambda$  now 
$\Lambda_2$), jumps to $- \infty$, and then increases to zero again, at
$\Lambda$ equal to $\Lambda_3$. For large $n$, we conclude that:
\beq
\Lambda_1^2 F_n(0) = \Lambda_3^2 F_{n+1} (0)\ ,
\eeq 
due to an intermediate matching:
\beq
\Lambda_1^2 F_n(0) = \Lambda_2^2 F_n (\infty ) = \Lambda_2^2 F_{n+1} 
(- \infty) = \Lambda_3^2 F_{n+1} (0)\ .
\eeq 
These equations establish that the ratio of successive bound state energies,
for large enough $n$, is equal to the square of the ratio $\Lambda_3 /
\Lambda_1$; this ratio is in turn the exponential of the period $t_2 - t_1$ of
the limit cycle:
\beq
{B_3^{(n)} \over B_3^{(n+1)}} = {\Lambda_3^2 \over \Lambda_1^2} 
    = \exp [ 2 (t_2 - t_1)]
\ ,
\eeq 
which is valid for all $n$ at the two-body fixed point with a
zero-energy bound state and valid in the limit of large $n$ when the
two-body binding energy is non-zero, as we will now discuss.


\section{Three-body Renormalization Group: A Second Look}
\label{3bdy_rg2}

        In this section, we consider the renormalization group
equations for the three-body sector in the presence of a non-zero
two-body energy $B_2$  and the two two-body couplings $G_2$ and  $h_2$.
We examine how $G_2$ and  $h_2$ affect $G_3 (\Lambda)$, when both are
near their fixed point values $G_2^*(h_2^*)$ and  $h_2^*$. We draw on
results from following sections, but only of a very general kind, in
order to arrive at somewhat surprising conclusions.

        The first conclusion justified in the following sections is
that the renormalization group equations for $G_2$ and $h_2$ apply
unchanged for the  three-body sector, and moreover,  there are no
further irrelevant operators which affect three-body energies to
relative order $1/\Lambda$. Further couplings need to be defined only
if one is interested in corrections of relative order 
$\ln\Lambda/\Lambda^2$, for example. Thus the only change we need to
make from our previous discussion is to generalize the renormalization
group equation for  $G_3( \Lambda)$ to include dependence on $G_2$ and 
$h_2$:
\beq
\Lambda {d G_3 \over d \Lambda} = \beta_{G_3} (G_2, G_3, h_2)\ .
\eeq 
If $G_2$ and  $h_2$ are both at their fixed points, then $G_3(\Lambda
)$ exhibits a limit cycle as before, which we denote as $g_{3c}
(\Lambda )$. The limit cycle complicates the analysis for the case that
$G_2$ and  $h_2$ are near but not at their fixed points. We can avoid
these complications by setting up a finite difference equation linking
the values of  $G_3(\Lambda )$ at a discrete set of values for
$\Lambda$  that are all separated by integer multiples of the limit
cycle period. To be specific, we let $\Lambda_0$  be an initial very
large cutoff, and then define  $\Lambda_1$,   $\Lambda_2$, etc. to be a
diminishing sequence of cutoffs separated by the limit cycle period:
\beq
\Lambda_{n+1} = \exp [ - (t_2 - t_1)]\Lambda_n\ .
\eeq 

        We now study departures from the exact limit cycle for $G_3(\Lambda )$
and exact fixed points for $G_2$ and $h_2$  in terms of small differences from
them:
\begin{eqnarray}
\delta g_{2_n} & = & G_2(\Lambda_n) - G_2^* (h_2^*),\\
\delta g_{3_n} & = & G_3(\Lambda_n) - g_{3c} (\Lambda_n),\\        
\delta h_{2_n} & = & h_2(\Lambda_n) - h_2^*.
\end{eqnarray}
To first order in these small quantities,
\beq
\delta g_{2_n}  = {a \sqrt {B_2} \over \Lambda_n}\ ,
\eeq 
where $a$ is a numerical constant and
\beq
\delta h_{2_n}  = {\delta h_{2_0} \Lambda_n \over \Lambda_0}\ .
\eeq 
This last equation comes from Eq. (27).

For $\delta g_{3n}$   we have a recursion formula:
\beq
\delta g_{3_{n+1}} = \delta g_{3_n} + b_0 \delta g_{2_n} + b_2 \delta h_{2_n}
+b_{02} \delta g_{2_n}\delta h_{2_n} + \dots\ .
\eeq 
The recursion formula includes all further possible polynomial terms in
the three variables $\delta g_{2_n}$, $\delta g_{3_n}$, and  $\delta
h_{2_n}$, except that no further powers of $\delta g_{3_n}$ by itself
can appear: such terms would be contrary to the existence of the limit
cycle when $\delta g_{2_n}$ and $\delta h_{2_n}$ are both zero.

        The recursion formula is easily solved by iteration, to determine
$\delta g_{3_n}$ in terms of  $\delta g_{3_0}$, $B_2$, $\delta
h_{2_0}$, and the ratio  $\Lambda_n / \Lambda_0$. Of special interest
is that in addition to powers of  $\Lambda_n / \Lambda_0$, a logarithm
appears too, in the form of a factor of $n$. The factor of $n$ is
generated from the $b_{02}$ term in the recursion formula, as can
easily be verified. As a result, $\delta g_{3_n}$ is
\beq
\delta g_{3_n} = \delta g_{3_0} + {(na) (\delta h_{2_0}) \sqrt {B_2} \over
\Lambda_0} + \dots\ .
\eeq 
This simple analysis leads us to expect that if we use a single
three-body interaction, $B_3$ will display residual cutoff dependence
of order $B_2 {\rm ln}(\Lambda)/\Lambda^2$.

Finally, we establish over the next few sections that the eigenvalues
$B_3^{(n)}$, for large enough $n$, are given in terms of a universal
function $F$:
\beq
B_3^{(n)} = \Lambda_n^2 F (g_{2_n}, g_{3_n}, h_{2_n})
\eeq 
%
%
with an error of order  $\ln\Lambda_0/\Lambda_0^2$. This result enables the
use of ``effective field theory" methods applied to the three-body case, with
errors of order   $\ln\Lambda^\prime/(\Lambda^\prime)^2$ where
$\Lambda^\prime$  is the cutoff used in the effective Hamiltonian.


\section{Integral Equations for a Cutoff Three-Body System}
\label{3bdy_inteq}

With an exact renormalization of the two-body problem, we can continue
to the three-body problem. The two-body interaction, with couplings
$G_2$ and $h_2$, is determined and we will see that a three-body
interaction is required to obtain a $\Lambda \rightarrow \infty$ limit
for the three-body problem \cite{BHK99,BHK99b,BHK00}. Once again, we
are interested in first obtaining the continuum limit and then
controlling cutoff dependence using an expansion in inverse powers of
the cutoff so that this model can be used as an effective field theory
in both the two-body and the three-body sectors.

One difference between our calculations and that of Bedaque et al. is
that we consistently cut off both the two-body and three-body
interactions using Gaussian cutoffs. We believe that this is ultimately
necessary if we want to tune the theory away from the continuum limit,
because as we have seen the continuum limit will not typically
correspond even at low energies to the theory we wish to model ({\it
e.g.}, QED or QCD). The same procedure was used in a recent extension
of effective field theory methods to the four-body problem by Platter
et al \cite{Platter04}.

A Gaussian cutoff is chosen because it enables us to use very accurate
numerical methods. In particular, we are able to obtain exponential
convergence for a uniform logarithmic scale \cite{Tadmor} in the finite difference
equations we ultimately solve, as is explained in detail in Mohr's
thesis \cite{Mohr03}. We maintain 10-12 digits of precision over about
50-70 orders of magnitude of momentum when solving the bound state
integral equation numerically.

In position space, the three-body bound-state equation takes the form
\begin{eqnarray}
- B_3 \, \psi(\vec{r_1}, \vec{r_2}, \vec{r_3}) & = & \left[ -\nabla_1^2 
   - \nabla_2^2 - \nabla_3^2 \right. \nonumber \\
 & & - \, g_2 \, \delta^3(\vec{r_1} - \vec{r_2}) - g_2 \, 
   \delta^3(\vec{r_2} - \vec{r_3}) - g_2 \, \delta^3(\vec{r_3} 
   - \vec{r_1}) \nonumber \\
 & & \left. + \, g_3 \, \delta^3(\vec{r_1} - \vec{r_2}) \, 
    \delta^3(\vec{r_2} - \vec{r_3}) \right] \, 
    \psi(\vec{r_1}, \vec{r_2}, \vec{r_3}) 
 \label{eqn:pos3bodyham}.
\end{eqnarray}
\noindent Here, $B_3$ is the three-body bound-state energy, and $g_3$
is the dimensionful coupling strength of a three-body contact
interaction which acts only when all three particles are at the same
point.  We will exchange the dimensionful couplings $g_2$ and $g_3$ for
the dimensionless couplings $G_2$ and $G_3$ when we regulate the
interaction. The form of the three-body interaction is somewhat
arbitrary.  We have chosen the product of two delta functions because
it is the simplest one that is non-zero only when $\vec{r_1} =
\vec{r_2} = \vec{r_3}$.  Other forms could be used, but the results
will remain unchanged if universality holds.

In order to simplify the three-body Schr{\" o}dinger equation in
momentum space we can assume without loss of generality that ${\vec
p_1} + {\vec p_2} + {\vec p_3} = 0$, and we can use the symmetry of the
wave function under exchange of coordinates, $\phi({\vec p_1} ,{\vec
p_2} ,{\vec p_3}) = \phi({\vec p_1} ,{\vec p_3} ,{\vec p_2})$, etc.

To regulate we again replace delta functions by
cutoff functions of the form
\beq
{\widetilde U}_\Lambda (\vec{r}_1 - \vec{r}_2) = \int d \tau \exp [ i \vec{p} \cdot 
(\vec{r}_1 - \vec{r}_2)] U_\Lambda(\vec{p})\ .
\eeq 
The cutoff form of the two-body potential 
is:
\beq V(\vec{r}_1, \vec{r}_2, \vec{r}_1^\prime, \vec{r}_2^\prime)=
-g_2 \widetilde{U}_\Lambda (\vec{r}_1 - \vec{r}_2) \widetilde{U}_\Lambda
(\vec{r}^\prime_1 - \vec{r}^\prime_2) \delta^3 \left [ 
{ \vec{r}_1 + \vec{r}_2 - (\vec{r}_1^\prime + \vec{r}_2^\prime) \over 2} 
\right ] \ .
\eeq
In general we need to use different regulator functions, $U_2$ and
$U_3$, in the two-body and three-body interactions in order to
introduce separate irrelevant operators in each sector. For example,
this allows us to use $h_2$ as above and introduce $h_3$ in the
three-body sector. For notational simplicity we will usually drop the
$\Lambda$ subscript since these couplings are always cutoff dependent.
         
We omit details of all the transformations to momentum space (see Ref.
\cite{Mohr03}), instead turning directly to the Schr\"odinger equation
in momentum space for an energy eigenvalue $-B_3$. In the
center-of-mass frame the Schr\"odinger equation is:
\begin{eqnarray}
- B_3 \varphi (\vec{p}_1, \vec{p}_2) & = & 
(\vec{p}_1^2 + \vec{p}_2^2 + \vec{p}_3^2) \varphi (\vec{p}_1, \vec{p}_2) 
- g_2 U_2(\vec{p}_1 + \vec{p}_2 / 2) \Phi (\vec{p}_2) \\ \nonumber
&& - g_2 U_2(\vec{p}_2 + \vec{p}_1 /2) \Phi (\vec{p}_1) 
-g_2 U_2(\vec{p}_1 / 2 - \vec{p}_2 / 2 ) \Phi (\vec{p}_3) \\ \nonumber
&& + g_3 U_3 (\vec{p}_1) U_3(\vec{p}_2) U_3(\vec{p}_3) \Phi_1\ ,
\end{eqnarray} 
where
\begin{eqnarray}
\Phi (\vec{p}) &=& {1 \over (2 \pi)^3} \int d^3 q U_2(\vec{q} + \vec{p}/ 2) 
\varphi (\vec{q}, \vec{p})\ ,\\
\Phi_1 & = & {1 \over (2 \pi)^6} \int d^3 {\vec p}_1 \int d^3 {\vec p}_2 
U_3 (\vec{p}_1) U_3(\vec{p}_2) U_3(\vec{p}_3) \varphi
(\vec{p}_1, \vec{p}_2)\ ,
\end{eqnarray} 
and
\beq
g_3 = {G_3(\Lambda) \over \Lambda^4}\ .
\eeq
In the following we further restrict
$\varphi$ to be in an S-wave of the total angular momentum. This
means that  $\Phi (\vec{p})$ depends only on the scalar variable $p$, where 
$p^2$ is  $\vec{p}^2$.

The Schr\"odinger equation allows $\varphi(\vec{p}_1, \vec{p}_2)$ to be
expressed in terms of  $B_3$, the function  $\Phi (p)$ and  $\Phi_1$.
In consequence, one can derive an integral equation for $\Phi (p)$ 
which has to be solved simultaneously with an algebraic equation for 
$\Phi_1$. The complete set of bound state equations  is:
\begin{eqnarray}
\Phi(p) & = & \frac{2}{D\left(-\frac{3}{2}p^2-B_3\right)} \int_0^{\infty} 
   \frac{q^2 dq}{4 \pi^2} \int_{-1}^{1} dz \: 
   \frac{U_2(\vec{q}+\frac{1}{2}\vec{p}\,) U_2(\vec{p}+\frac{1}{2}\vec{q}\,)}
      {2p^2 + 2q^2 + 2pqz + B_3} \Phi(q) \nonumber
\\
&& ~~~~~~~- \frac{g_3 \, D_1(p) \, \Phi_1}{g_2 \, 
    D\left(-\frac{3}{2}p^2-B_3\right)} ,
\label{phieq}
\\
D(E) & = & \frac{1}{g_2} - {\cal P} \int \frac{d^3\vec{q}}{(2 \pi)^3} \, 
   \frac{U_2(q)^2}{2q^2 - E} ,
\label{Deq}
\\
D_1(p) & = & \frac{1}{4 \pi^2} \int_0^{\infty} q^2 dq \int_{-1}^{1} dz \: 
  \frac{U_2(\vec{q}+\frac{1}{2}\vec{p}\,) U_3(q) U_3(p) U_3(\vec{p}+\vec{q}\,)}
     {2p^2 + 2q^2 + 2pqz + B_3},
\label{D1eq}
\\
\Phi_1 & = & \frac{3 g_2}{2 \pi^2} \int_0^{\infty} dq \left[ q^2 D_1(q) 
     \Phi(q) \right] - g_3 \, D_2 \Phi_1 ,
\\
D_2 & = & \frac{1}{8 \pi^4} \int_0^{\infty} p^2 dp \int_0^{\infty} q^2 dq 
   \int_{-1}^{1} dz \: \frac{U_3(q)^2 U_3(p)^2 U_3(\vec{q}+\vec{p}\,)^2}
      {2p^2 + 2q^2 + 2pqz + B_3} ,
\label{D2eq}
\end{eqnarray}
where ${\vec p} \cdot {\vec q} = p q z$.
The quantities  $\Phi(p)$,  $\Phi_1$,  $D_1$, and  $D_2$ all depend on
$B_3$  and  $\Lambda$, although this has not been indicated explicitly.
The quantities  $\Phi(p)$,  $\Phi_1$, and $D$ all depend on $B_2$  as
well, and $D$ depends implicitly also on  $\Lambda$.

If  $G_3$ is zero, only the equation for $\Phi(p)$  matters, and it
will have solutions only when  $B_3$ is at an eigenvalue. If $G_3$  is
non-zero, then one can generate a solution for any value of  $B_3$,
namely by specifying a value for the product   $G_3 \Phi_1$. This
specification turns the equation for  $\Phi(p)$ into an inhomogeneous
equation that should pose no constraint on  $B_3$. Once the equation
for $\Phi(p)$  has been solved, the equation for $\Phi_1$  can be used
to determine $\Phi_1$   and hence  $G_3$ as well. We will see that if
we fix $G_3(\Lambda)$ to remove cutoff dependence for one eigenvalue,
so that the $\Lambda \rightarrow \infty$ limit can be explicitly taken,
the limit for a complete tower of eigenvalues will be determined.

In order to precisely solve this set of equations we introduce a method
that allows us to disentangle small, intermediate and large momentum
regions. This method is an analog of Efimov's disentanglement of short,
intermediate and long distances in position representation
\cite{Efi70,Efi71}.


\section {Uniformly Valid Expansion for a Simple Function}
\label{sol_examp}

The next problem we address is the determination of the limiting
behavior of the solution  $\Phi(p)$ of the integral equation for very
large  $\Lambda$, including subdominant terms that are smaller by a
factor of $\Lambda$ or  $\Lambda^2$ than the leading term. We want to
first fix the bound state equation for the infinite cutoff limit and
then develop power series expansions in inverse powers of the cutoff
for the eigenvalue and wave function.

We want to solve an equation involving small binding energies in
comparison to the cutoff, because we want to start in the infinite
cutoff limit. It will turn out that we can derive the leading order
equations analytically and show that they reduce to an equation for the
low-momentum part of the wave functions that must match at intermediate
momenta a function that we derive analytically that depends on a single
angular variable. We have a small three-body binding energy $B_3$ and
we will allow a small two-body binding energy $B_2$, so $B/\Lambda^2$
is a small quantity. But the momentum in the bound-state equation
ranges from small values all the way up to infinity, with the wave
function dying exponentially above the cutoff.

The strategy we use is to divide the range of $p$ into three parts: $0
< p \simlt \sqrt{B_3},\,  \sqrt{B_3} \ll p \ll \Lambda$, and  $\Lambda
\simlt p < \infty$.  We use separate numerical computations to handle
the lowest and highest of these ranges, while solving the middle range
analytically.

        There is a formal analysis that underlies our breakdown of the
problem into the three parts. We will introduce the formal analysis
with a far simpler problem. We consider a simple function of three
variables:  $\eta$, $p$, and  $\Lambda$. The function we examine is:
\beq
f (\eta, p, \Lambda) = {1 \over (p + \eta) (p + \Lambda)}\ .
\eeq
We will offer a uniformly valid approximation scheme for $f$, valid
over the whole range of $p$. The expansion converges as a power series
in $\eta/\Lambda$, but we cannot naively expand because $p$ ranges from
order $\eta$ to infinity. We will then discuss how to compute integrals
over the full range of  $p, 0 < p < \infty$, using this scheme. 

The formula is built from three expansions for $f$ that are valid only
in limited domains of $p$. Each of the limited expansions is valid in
one of the three ranges for $p$ already mentioned. We assume $\eta \ll
\Lambda$, so in every domain we can expand in powers of $\eta/\Lambda$.
In addition, when $p \simlt \eta \ll \Lambda$, we can also expand in
powers of $p/\Lambda$. We refer to this expansion as  $f_l (\eta, p,
\Lambda)$. In the highest range for $p$, $p \simgt  \Lambda$, we can
expand instead in $\eta /p$; we refer to this second expansion as 
$f_h(\eta, p, \Lambda)$. In the middle range   $\eta\ll p \ll \Lambda$,
we can use a double expansion in powers of both $p/\Lambda$  and $\eta
/p$; we call this double expansion  $f_d (\eta, p, \Lambda)$. Our
uniformly valid formula is
\beq
f(\eta, p, \Lambda) = f_l(\eta, p, \Lambda) + f_h(\eta, p, \Lambda) 
- f_d(\eta, p, \Lambda) \ .
\eeq
The simple, intuitive way to understand how this formula allows us to
produce an approximation for the full function that is valid to any
given order in $\eta/\Lambda$ in every region of momentum is to observe
that $f_d$ will equal $f_l$ when $\eta \ll p$ and it will equal $f_h$
when $p \ll \Lambda$. Thus in these regions, where $f_l$ and $f_h$ are
not valid, they are exactly cancelled by $f_d$ to whatever order we
choose to work. In the intermediate region, both $f_l$ and $f_h$ are
valid, so $f_d$ simply cancels a double-counting in this region.

        Suppose, for example, that one wants an approximation for $f$ that is
uniformly valid apart from errors of third order in the ratio $\eta /\Lambda$.
Then the rule is that we first construct truncated versions of $f_l (\eta, p,
\Lambda)$  and  $f_h (\eta, p, \Lambda)$, valid to this accuracy in the
corresponding range for $p$. Thus, for $p\sim \eta$, we expand $f$ through
order $(p/\Lambda)^2$, neglecting order  $(p/ \Lambda)^3$. The resulting
truncated expansion is:
\beq
f_l(\eta, p, \Lambda) = {1 \over (p + \eta) \Lambda} - {p \over (p + \eta)
\Lambda^2} + {p^2 \over (p + \eta) \Lambda^3}\ .
\eeq
Similarly, we expand $f$ through order $(\eta / p)^2$  to give the truncated
expansion for large $p$:
\beq
f_h(\eta, p, \Lambda) = {1 \over p (p + \Lambda)} - {\eta \over p^2 (p +
\Lambda)} + {\eta^2 \over p^3 (p + \Lambda)}\ .
\eeq
Now we come to the crucial rule: to determine the truncated form of the double
expansion $f_d (\eta, p, \Lambda)$, one first constructs the doubly expanded
versions of $f_l (\eta, p, \Lambda)$  and  $f_h (\eta, p, \Lambda)$. One starts
from their truncated forms just given, and in the double expansions, one keeps
all terms that are necessary to preserve the accuracy of the double expansion
through relative order $(\eta / \Lambda)^2$  at the {\it opposite end} of the
range of $p$.  For example, the doubly expanded version of  $f_l (\eta, p,
\Lambda)$, which we can denote by  $f_{ld} (\eta, p, \Lambda)$, is computed by
expanding  $f_l (\eta, p, \Lambda)$. We assume that $p$ is of order  $\Lambda$,
and keep terms through relative order  $(\eta / \Lambda)^2$:
\beq
f_{ld}(\eta, p, \Lambda) = \biggl( {1 \over p \Lambda}
   - {\eta \over p^2 \Lambda}
   +{\eta^2 \over p^3 \Lambda}\biggr)
   - \biggl({1 \over \Lambda^2}-{\eta \over p \Lambda^2}
   +{\eta^2 \over p^2\Lambda^2}\biggr)
   +\biggl({p \over  \Lambda^3}-{\eta \over  \Lambda^3}
   +{\eta^2 \over p \Lambda^3}\biggr)\ .
\eeq
We likewise expand $f_h(\eta, p, \Lambda)$  through relative order $(\eta /
\Lambda)^2$  assuming that $p$ is of order  $\eta$, giving:
\beq
f_{hd}(\eta, p, \Lambda) = \biggl({1 \over p \Lambda}- {1 \over \Lambda^2}
  +{p \over\Lambda^3}\biggr) - \biggl({\eta \over p^2 \Lambda}
  -{\eta \over p \Lambda^2}
  +{\eta \over\Lambda^3}\biggr)+ \biggl({\eta^2 \over  \Lambda p^3}
  -{\eta^2 \over  
  p^2 \Lambda^2}+{\eta^2 \over p \Lambda^3}\biggr)\ .
\eeq
We note that the two double expansions are identical to each other,
apart from a rearrangement of terms. We believe that this is always the
case when applying the rules stated above, although we offer no proof
of this claim. Now the rule for truncating the complete double
expansion  $f_d(\eta, p, \Lambda)$ is to set it equal to these
identical double truncations:
\beq
f_d(\eta, p, \Lambda) = f_{ld}(\eta, p, \Lambda) = f_{hd}(\eta, p, \Lambda)
\ .
\eeq
Given these definitions for truncated expansions to include in the formula for
$f$, does the formula work?

\begin{figure}
\begin{center}
\includegraphics[width=5.0in]{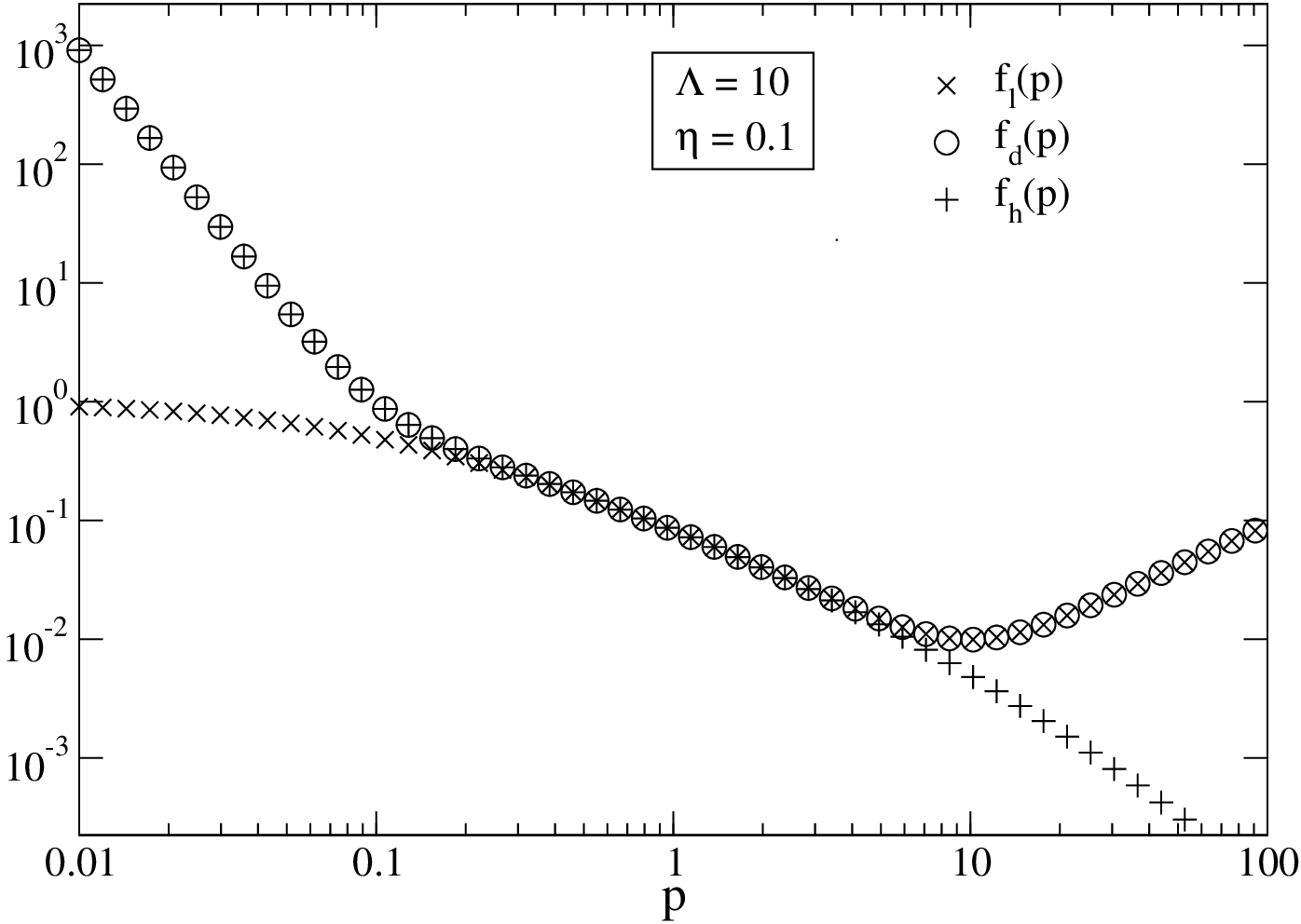}
\end{center}
\vspace*{-.1in}
\caption{Components of the uniformly valid
expansion of a simple function. Note that $f=f_l+f_h-f_d$. For small
momenta one clearly sees $f_h$ and $f_d$ will cancel, for high momenta
$f_l$ and $f_d$ will cancel, and in the middle one set will cancel,
leaving $f$ uniformly approximated in all regions.}
\label{fig:simple_function} 
\end{figure}

        To show that the formula works, one has to consider the three
ranges of $p$ separately. In the low range: $p \simlt \eta$,  where $f$
is approximated by  $f_l (\eta, p, \Lambda)$, the other two terms
cancel to the required accuracy. In the high range, $p \simgt 
\Lambda$, $f$ is approximated by $f_h (\eta, p, \Lambda)$  and the
other two terms cancel to the required accuracy. Thus the key region to
discuss is the intermediate range  $\eta \ll p \ll\Lambda$. What
happens in the intermediate range is that both $f_l (\eta, p,
\Lambda)$  and $f_h (\eta, p, \Lambda)$  have double expansions in 
$\eta/p$ and $p/ \Lambda$. But $f_l (\eta, p, \Lambda)$  has terms to
all orders in $\eta /p$ while only to second order in $p/\Lambda$. The
function $f_h (\eta, p, \Lambda)$ has the reverse situation -- an
expansion to all orders in $p/\Lambda$  but only to second order in
$\eta /p$. When the two double expansions are added together, the terms
that are of second order or less in both expansion variables are double
counted. The subtraction of  $f_d (\eta, p, \Lambda)$ removes the
doubly counted terms. As a result the approximation for $f$ is in error
due only to terms which are of at least third order in both expansion
parameters, and the error caused by a product of third order terms is
\beq
(\eta/p)^3 (p/\Lambda)^3 = (\eta /\Lambda)^3\ .
\eeq
For this simple example, the exact error can be determined, as a check on the
discussion above. We offer two intermediate results followed by the error
formula:
\begin{eqnarray}
f(\eta, p, \Lambda) - f_l(\eta, p, \Lambda) &=& {-p^3 \over \Lambda^3 (p +
\eta) ( p + \Lambda)} \ ,\\
f_h(\eta, p, \Lambda) -f_d(\eta, p, \Lambda) &=& {-p^3 \over \Lambda^3 (p +
\Lambda)} \left \{ {1 \over (p + \eta)} + {\eta^3 \over p^3 (p+ \eta)} \right
\}\ .
\end{eqnarray}
The error in the approximation for $f$ is the difference of these two
expressions, namely:
\beq
{\eta^3 \over \Lambda^3 (p + \eta) ( p + \Lambda)}\ .
\eeq
{\it This error is smaller than $f$ by a factor of  $(\eta /
\Lambda)^3$ for all momenta}, as promised.
Figure~\ref{fig:simple_function} shows the three components of the
expansion, $f_l$, $f_d$ and $f_h$. The exact function is given by
$f_l+f_h-f_d$ and at this level of approximation for these components,
the difference between the exact function and the approximation can not
be seen in such a figure.

To complete our discussion of our simple function $f$, we discuss a
problem that arises when integrating the uniformly valid approximation
over the whole range $0 < p < \infty$. We assume that the integral of
the exact functions converges, so clearly if we use the entire
uniformly valid expansion to approximate the function the integral will
converge, producing an expansion in powers of $\eta/\Lambda$ of the
exact integral. The problem arises only if this integration is
performed term by term; in this case some of the terms can lead to
divergent integrals. For one example, the two most divergent terms for
large $p$ are:
\beq
{p^2 \over (p + \eta) \Lambda^3} - {p \over \Lambda^3}\ ,
\eeq
each of which give rise to quadratically divergent integrals for $p$
near infinity. The first term also has subdominant linear and
logarithmic divergences when it is expanded in powers of  $\Lambda /
p$. All these divergences are cancelled: the quadratic divergences
cancel between these two terms, and the subdominant divergences are
cancelled by other terms from  $f_d$. 

The divergences must cancel exactly, but it is convenient for analytic
work to have a procedure that  makes the integral of each term
separately finite, without changing the integral of the sum of all
terms. We can accomplish this goal by introducing a standardized
subtraction to apply to each divergent integral to make it finite. For
example, we could suggest the following set of standardized
subtractions to make all integrals finite at their upper limit:

\begin{enumerate}
\item  Positive powers of $p$ and constant terms independent of $p$
are dropped completely, e.g., $p^n$  is replaced by 0 for all $n\geq 0$.
\item A logarithmic divergent $1/p$ term, integrated to infinity, is 
subtracted out from $\Lambda$  to infinity. 
\end{enumerate}

According to these rules the $p /\Lambda^3$  term would be dropped
completely, positive powers of $p$ for $p \gg \Lambda$    would be
subtracted from the $p^2 / \left ( (p+ \eta) \Lambda^2 \right )$  term,
and the $1/p$ component of this term would be subtracted out only for
$p >  \Lambda$. To illustrate how this rule works, we exhibit the
subtracted form for the integral over the first term:
\beq
  \int^\infty_0 dp \left \{ {p^2 \over (p + \eta) \Lambda^2} - {p \over
  \Lambda^2} + {\eta \over \Lambda^2} \right \} - \int^\infty_\Lambda {\eta^2
  \over p \Lambda^2} dp\ .
\eeq
Note that a rule subtracting $1/p$ terms for all $p$ would not work, because
they would introduce unwanted divergences for $p\to 0$.
These rules do not change the overall integration of the uniformly valid
approximation because all the subtractions cancel out.

A similar set of rules can be derived for handling divergences for 
$p \to 0$:

\begin{enumerate}
\item Negative powers of $p$ other than $1/p$ are dropped completely.
\item A logarithmically divergent $1/p$ term is subtracted out only for 
$0 < p < \eta$. 
We use $\eta$  rather than $\Lambda$  because $1/p$ divergences for small
$p$ typically emerge from functions involving  $\eta$ rather than  $\Lambda$,
and this form for the subtraction ensures that such functions do not acquire an
unwanted dependence on $\Lambda$  as well.
\end{enumerate}

The application of this scheme is easy when given a simple function,
but its real power is only seen when solving an integral equation in
which the function is not known. Even though the function is not known,
we can solve for the pieces of its uniformly valid expansion and this
is our procedure for obtaining an expansion of $B_3$ and the wave
function in powers of $B/\Lambda^2$.

In order to solve the complete set of equations, we also need an
expansion for functions that depend on two momenta, $p$ and $q$, each
of which range from $0$ to $\infty$. We do not present a full analysis
of this expansion here, because it requires eleven regions. A full
discussion is found in Mohr's thesis \cite{Mohr03}.


\section {Leading Order Three-body Equation for Large  $\Lambda$}
\label{sec:sol_inteq}

We are now prepared to solve the three-body equations in the limit
$\Lambda \rightarrow \infty$. To do this, we first need to find the
leading-order approximation to the full integral equations using the
method of uniformly valid expansions. We assume that all binding
energies are much less than the cutoff and isolate approximations that
are valid in various regions of momenta. It is conceptually
straightforward to extend the calculation to systematically include
higher-order corrections, but these will not affect the continuum limit
that we isolate as a starting point. We will provide only those details
that are required to qualitatively understand the results and once
again refer the reader to Mohr's thesis \cite{Mohr03} for details.

We start with a few definitions that will simplify notation somewhat:
\begin{eqnarray}
\eta_2 & \equiv & \sqrt{B_2} ,
\\
\eta_3 & \equiv & \sqrt{B_3} .
\end{eqnarray}
\noindent Using $\eta_2$ and $\eta_3$ in our equations allows us to
deal with quantities that have the same dimension as the momentum
variable.  It also identifies their role in the expansion as the
parameter $\eta$ from the previous chapter.  In fact, we will
occasionally use $\eta$ to generically refer to $\eta_2$ and $\eta_3$
in places where either is a valid alternative.

We use dimensionless couplings, $G_2  \equiv  \Lambda \, g_2$ and $G_3 
\equiv  \Lambda^4 \, g_3 .$ Further, we note that $G_3$ enters the
bound-state equation only as a multiplicative constant that is easier
to work with:
\begin{equation}
\delta \equiv \frac{G_3 \, \Phi_1}{\Lambda} .
\end{equation}
$\delta$ will be found to remain finite for all values of the cutoff,
and at the end of the calculation we can readily compute $\Phi_1$ to
isolate $G_3$ and display the limit cycle. We will find that $\Phi_1$
is independent of $B_3$ to leading order, which is consistent with the
fact that we have only one coupling $G_3$ with which we must
renormalize the entire bound state spectrum.

Another dimensionless quantity, and perhaps the most important, is the
replacement of the pseudo-wavefunction $\Phi(p)$ with
\begin{equation}
f(\eta_2, \eta_3, p, \Lambda) \equiv (p^2 + \eta_3^2 - \eta_2^2) \Phi(p).
\end{equation}
\noindent  Like $\eta_2$ and $\eta_3$, its role in the expansion is
easily seen to be the same as the function $f$ used in the previous
chapter.  Like $G_2$ and $G_3$, the dimensionless nature of $f$ will
simplify future power counting.  In addition, the $p^2$ term in the
factor $p^2+\eta_3^2-\eta_2^2$ ensures that $f$ tends to be of ${\cal
O}(1)$ throughout the entire range of $p$ while the $\eta_3^2-\eta_2^2$
term makes $f$ less prone to large numerical fluctuations than $\Phi$
when $B_2 \simeq B_3$.\footnote{We cannot prove these statements  {\em
a priori} and must verify them numerically.}

With these new definitions, the three-body bound-state equation
(\ref{phieq}) now takes the form
\begin{eqnarray}
f(\eta_2, \eta_3, p, \Lambda) & = & 
   \frac{p^2 + \eta_3^2 - \eta_2^2}
        {2 \pi^2 D\left(-\eta_3^2 - \frac{3}{2}p^2\right)} 
           \int_0^{\infty} dq \: \left[ \frac{q^2}{q^2 + \eta_3^2 - \eta_2^2} 
           \right. \nonumber
\\
&& \left. \times \int_{-1}^{1} dz \: 
   \frac{U_2\left(\vec{q} + \frac{1}{2}\vec{p}\right) U_2\left(\vec{p} 
       + \frac{1}{2}\vec{q}\right)}{\eta_3^2 + 2p^2 + 2q^2 + 2pqz}  
       f(\eta_2, \eta_3, q, \Lambda) \right] \nonumber
\\
&& - ~\delta \; \frac{(p^2 + \eta_3^2 - \eta_2^2) D_1(p)}
                {G_2 \Lambda^2 D\left( -\eta_3^2 - \frac{3}{2}p^2 \right)} 
    \label{feq}.
\end{eqnarray}

The process of expanding Eq.~(\ref{feq}) is done in a series of steps. 
Because the $\delta$ term is simply added to the integral term, each
can be expanded individually and then added together at the end.  In
addition, these terms are composed of other quantities like $D_1(p)$,
$G_2$, etc. which have their own expansions in terms of
$\eta/\Lambda$.  The goal is to split the equation for $f$ into
separate equations for $f_l$, $f_d$, and $f_h$, isolating and solving
the leading order equations to produce a uniformly valid approximation
for the wave function. The details of the calculation are not very
enlightening\footnote{See chapter 5 of Ref. \cite{Mohr03}.}, so we
simply list the complete set of leading order equations that must be
solved.

\bigskip

\noindent {\bfseries Pseudo-Wavefunctions:}

We are able to precisely compute a cutoff integral over one variable of
the three-body wave function. The fundamental equations for the
calculation of these pseudo-wavefunctions are:
\begin{eqnarray}
 f_{l0}(\eta_2, \eta_3, p) & = & \frac{1}{4 \pi^2 D_{l0}(\eta_2,\eta_3,p)} 
 \int_0^{\infty} dq \: \left[ 
    \frac{q \, (p^2 + \eta_3^2 - \eta_2^2)}{p \, (q^2 + \eta_3^2 - \eta_2^2)} 
                        \right.  \nonumber
\\
 && \times \left. \ln\left(\frac{\eta_3^2 + 2p^2 + 2q^2 + 2pq}
      {\eta_3^2 + 2p^2 + 2q^2 - 2pq}\right) f_{l0}(\eta_2,\eta_3,q) 
      \right] ,\label{eqn:QRfl0}
\\
&&~\nonumber
\\\
f_{d0}(p) & = &\frac{p}{4 \pi^2 D_{d0}(p)} \int_0^{\infty} \frac{dq}{q} \, 
    \ln\left(\frac{p^2 + q^2 + pq}{p^2 + q^2 - pq}\right) f_{d0}(q),
\\
&&~\nonumber
\\
f_{h0}(p, \Lambda) & = & \frac{p^2}{2 \pi^2 D_{h0}(p,\Lambda)} 
    \int_0^{\infty} dq \, \int_{-1}^{1} dz \, 
    \frac{U_2\left(\vec{q} + \frac{1}{2}\vec{p}\right) U_2\left(\vec{p} 
      + \frac{1}{2}\vec{q}\right)}{2p^2 + 2q^2 + 2pqz} f_{h0}(q, \Lambda), 
    \nonumber
\\
 && \hspace{0.5in} -~\delta_0 \, \frac{3 h_2^2 + 8 h_2 + 16}
                 {128 \sqrt{2} \, \pi^{3/2}} 
                 \left( \frac{D_{1h0}(p,\Lambda)}{D_{h0}(p,\Lambda)} \right) 
                 \left( \frac{p^2}{\Lambda^2} \right).
\\
&&~\nonumber
\end{eqnarray}
The intermediate range can be determined analytically \cite{Dan61}:
\begin{eqnarray}
f_{d0}(p) & = & A \cos\left( s_0 \ln\left(\frac{p}{\Lambda}\right) 
    + \theta \right),
\\
&&~\nonumber
\end{eqnarray}
where we can choose $A=1$ because the bound state equations are
homogeneous, $\theta$ is a phase determined by boundary conditions, and
$s_0$ is the real, positive solution to the equation
\begin{equation}
\frac{8}{\sqrt{3}} \, \sinh\left(\frac{\pi s_0}{6}\right) = s_0 \, 
   \cosh\left(\frac{\pi s_0}{2}\right) .
\end{equation}
\noindent The value of $s_0$ to the precision we want is $1.006237825102$.

The $\Lambda$ dependence in Eq.~(85) may be a bit misleading
since Eq.~(83) does not contain it.  The cutoff is required because the
argument of the logarithm is dimensionless.  Making this choice is a
matter of preference, and it could just as easily have been $\eta_3$.
We discuss this renormalization prescription below; but first we need
to show how choosing two binding energies uniquely determines one and
only one bound state spectrum.

Clearly the choice of $B_2$ can be made independently of any modeling
in the three-body system, so we take this and $B_3$ as input for
Eq.~(85). We assume that $B_3 > B_2$ so that a stable three-body bound
state exists. This determines $f_{l0}(p)$, which must map exactly on to
$f_{d0}(p)$ for $p \gg \eta_3$. Once $\Lambda$ is chosen, $\theta$ in
$f_{d0}$ is determined by the phase of $f_{l0}$, which is in turn
determined by both $B_2$ and $B_3$. Finally, $f_{h0}(p)$ must map onto
$f_{d0}(p)$ for $p \ll \Lambda$. The solution requires a phase and this
fixes it. Given the phase in $f_{h0}(p)$, $\Phi_1$ is determined, which
in turn yields $G_3$.

This explains how one three-body bound state is determined. The
remaining allowed values of $B_3$ must produce the same phase $\theta$
as is produced by the value that is fixed to determine $\theta$. This
produces both the remaining spectrum and the pseudo-wavefunction.

The remaining equations we require are:

\bigskip

\begin{eqnarray}
   D_{l0}(\eta_2, \eta_3, p) & = & 
   \frac{\sqrt{\frac{3}{2}p^2 + \eta_3^2} - \eta_2}{8 \sqrt{2} \, \pi},
\\
&& \nonumber
\\
D_{d0}(p) & = & \frac{\sqrt{3} \, p}{16 \pi},
\\
&& \nonumber
\\
D_{h0}(p, \Lambda) & = & \frac{p}{256 \pi^2 \Lambda^4} 
   \left[ \sqrt{3} \, \pi \left( 4 \Lambda^2 - 3 h_2 p^2 \right)^2 
   \left( 1 - \mathrm{Erf}\left( \frac{\sqrt{3} \, p}{\sqrt{2} \, \Lambda} 
       \right) \right) 
   \exp\left( \frac{3 p^2}{2 \Lambda^2} \right) \right. \nonumber
\\
&& +~ 3 \sqrt{2 \pi} \, h_2 \Lambda p \left( \left(8 + h_2 \right) 
    \Lambda^2 - 3 h_2 p^2 \right) \Bigg],
\\
&& \nonumber
\end{eqnarray}

\begin{eqnarray}
D_{1l0}(\Lambda) & = & \frac{\Lambda \left( 6 \left( h_3^2 + 4 h_3 + 12 \right)
     + h_2 \left( 5 h_3^2 + 12 h_3 + 12 \right) \right)}
       {576 \sqrt{3} \, \pi^{3/2}},
\\
&& \nonumber
\\
D_{1d0}(\Lambda) & = & \frac{\Lambda \left( 6 \left( h_3^2 + 4 h_3 + 12 \right)
    + h_2 \left( 5 h_3^2 + 12 h_3 + 12 \right) \right)}
      {576 \sqrt{3} \, \pi^{3/2}},
\\
&& \nonumber
\\
D_{1h0}(p, \Lambda) & = & \frac{1}{4 \pi^2} \int_0^{\infty} dq \: q^2 \, 
   \int_{-1}^{1} dz \: \frac{U_2(\vec{q}+\frac{1}{2}\vec{p}\,) 
      U_3(q) U_3(p) U_3(\vec{p}+\vec{q}\,)}{2p^2 + 2q^2 + 2pqz},
\\
&& \nonumber
\end{eqnarray}

\begin{eqnarray}
G_2 & \equiv & \Lambda g_2,
\\
G_{2,0} & = & \frac{128 \sqrt{2} \, \pi^{3/2}}{3 h_2^2 + 8h_2 + 16},
\\
&& \nonumber
\end{eqnarray}

\begin{eqnarray}
G_3 & \equiv & \Lambda^4 g_3,
\\
G_{3,0} & = & \frac{\delta_0}{G_{2,0} \, {\cal I}_0 - D_{2,0}\,
   \delta_0/\Lambda^4 }, \label{eqn:QRG30}
\\
&& \nonumber
\end{eqnarray}

\begin{eqnarray}
{\cal I}_0 & = & \frac{3}{2 \pi^2 \Lambda^2} \int_0^{\infty} dp \, 
   D_{1h0}(p,\Lambda) \, f_{h0}(p,\Lambda),
\\
&& \nonumber
\end{eqnarray}

\begin{eqnarray}
D_{2,0} & = & \frac{1}{8 \pi^4} \int_0^{\infty} p^2 dp \int_0^{\infty} q^2 dq 
   \int_{-1}^{1} dz \: \frac{U_3(q)^2 U_3(p)^2 U_3(\vec{q}+\vec{p})^2}
                            {2p^2 + 2q^2 + 2pqz}.
\\
&& \nonumber
\end{eqnarray}

At this point we simply need to solve these equations numerically. We
demonstrate that we achieve 10--12 digits of precision.


\section {Discretization and Precision}
\label{sol}

Because a closed-form solution for either $f_{l0}$ or $f_{h0}$ is
unknown, we must resort to numerical calculations of these functions as
well as any energies or couplings.  A common method for solving an
integral equation involves changing the integration into a sum over
discrete points.  The integral equation then becomes a matrix equation
easily solved by standard methods.

While this may appear straightforward, there are several practical
issues to consider.  For example, new limits on the integral equation
must be determined.  It is impossible to numerically integrate to
infinity, so a suitable upper bound must be chosen.  In our case, we
use a logarithmic integration scale, which is critical in almost all
problems amenable to renormalization, so a new lower bound to replace
$0$ must also be determined.  How the discrete points are chosen must
be carefully considered.  Fortunately, for our problem {\it
exponential} convergence with the number of points can be achieved, as
discussed below.

Each choice is a compromise between accuracy and size.  For example, we
could choose an upper limit that is extremely large (ensuring that we
are ``close'' to infinity) and discrete points that are closely spaced
(minimizing errors in the sum).  The trade-off is that using more
points requires using a larger matrix.  Using $N$ discrete points will
result in an $N \times N$ matrix with $N^2$ elements.  If all numbers
are double precision decimals, even 8000 points would be enough to
overwhelm a computer with 512 MB of memory.  This does not even take
into account the time needed to process such a matrix.  Obviously, the
goal is to obtain the desired accuracy with a minimal number of
points.  We will discuss a few methods that drastically reduce the
number of points we need.

In the following sections, we will assume that our goal is about 12
digits of accuracy.  This high accuracy may not be necessary for most
leading order calculations, but it is essential when studying leading
corrections.  Besides directly obtaining the equations for the $\Lambda
\rightarrow \infty$ limit, one principal reason for expanding the
three-body equation in powers of $\eta/\Lambda$ is to analyze the
cutoff dependence.  If we are attempting to study this behavior, we
must be certain that our numerical errors are not larger than the
corrections being studied;  otherwise there is no way to distinguish
the small corrections from the numerical ``noise.''  

\subsection{Transition to Mid-Momentum Function}

One way to limit the size of the matrix is to limit the range over
which the function must be integrated.  We know that $f_{l0}$ and
$f_{h0}$ approach $f_{d0}$ for $p \gg \eta$ and $p \ll \Lambda$
respectively.  This allows us to replace either function with
$\cos\left(s_0 \ln(p/\Lambda) + \theta\right)$ in the appropriate
range.  The point at which we can make the switch is determined by the
accuracy we desire.  These limits are derived for the case of 12 digits
of accuracy.

For $p \gg \eta$, the equation for $f_{l0}$ can be written as
\begin{eqnarray}
&& \hspace{-0.25in} f_{l0}(\eta_2, \eta_3, p) = 
   \frac{4 (1 + \eta_3^2/p^2 - \eta_2^2/p^2)}
        {\sqrt{3} \pi \left( \sqrt{1 + (2 \eta_3^2)/(3 p^2)} 
          - (\sqrt{2} \eta_2)/(\sqrt{3} p) \right)} \int_0^{\infty} dq 
          \: \frac{q}{q^2 + \eta_3^2 - \eta_2^2} \nonumber
\\
&& \hspace{-0.25in} \times \left[ 
   \ln\left(\frac{p^2 + q^2 + pq}{p^2 + q^2 - pq}\right) + 
   \frac{\eta_3^2}{2 p^2 + 2 q^2 + 2 p q} 
   - \frac{\eta_3^2}{2 p^2 + 2 q^2 - 2 p q} \right] f_{l0}(\eta_2,\eta_3,q) .
\end{eqnarray}
\noindent Here we have treated $\eta/p$ as a small quantity and
perturbatively expanded all factors.  As long as both $\eta_2/p$ and
$\eta_3/p$ are less than $10^{-12}$, this equation will match the one
for $f_{d0}$ to 12 digits.  Of course, $\eta_3$ must be greater than
$\eta_2$ since we are considering only stable bound states.  This means
that $p \simeq 10^{12} \, \eta_3$ sets the limit above which $f_{l0}$
can be replaced by $f_{d0}$.  In practice, we must have enough data
points above this limit to ensure that our cosine fit is accurate to 12
digits also.  Therefore, we will use an actual limit of $p = 10^{15} \,
\eta_3$.

Similarly, we can expand the equation for $f_{h0}$ in the region $p \ll
\Lambda$.  Notice that in this region $D_{1h0}(p,\Lambda)$ approaches
$D_{1d0}(\Lambda)$ which is proportional to $\Lambda$, and the function
$D_{h0}(p,\Lambda)$ becomes equal to $D_{d0}(p) = \sqrt{3} \, p/16
\pi$.  In the $f_{h0}$ integral equation, the momentum-dependent part
of the three-body interaction becomes
\begin{equation}
  \left( \frac{p^2}{\Lambda^2} \right) \left( 
     \frac{D_{1h0}(p,\Lambda)}{D_{h0}(p,\Lambda)} \right) 
     \stackrel{p \ll \Lambda}{\longrightarrow} 
     \left( \frac{p^2}{\Lambda^2} \right) 
     \left( \frac{D_{1d0}(\Lambda)}{D_{d0}(p)} \right) 
     \propto \frac{p}{\Lambda} .
\end{equation}
\noindent Since the leading order mid-momentum equation has no
three-body interaction term, the above term will equal zero to 12
digits if we choose $p \sim 10^{-12} \, \Lambda$.

The integral part for $f_{h0}$ looks like
\begin{equation}
\frac{p^2}{2 \pi^2 D_{h0}(p,\Lambda)} \int_0^{\infty} dq \, 
  \int_{-1}^{1} dz \, 
  \frac{U_2\left(\vec{q} 
    + \frac{1}{2}\vec{p}\right) U_2\left(\vec{p} + \frac{1}{2}\vec{q}\right)}
        {2p^2 + 2q^2 + 2pqz} f_{h0}(q, \Lambda) .
\end{equation}
\noindent We have already stated that $D_{h0}(p,\Lambda)$ approaches
$D_{d0}(p)$, but more importantly, it approaches like
\begin{equation}
D_{h0}(p,\Lambda) \stackrel{p \ll \Lambda}{\longrightarrow} 
  D_{d0}(p) \left[ 1 + \left( 
    \frac{\sqrt{3} \, (h_2^2 + 8 h_2 - 16)}{8 \sqrt{2 \pi}} \right) 
    \frac{p}{\Lambda} \right] .
\end{equation}
\noindent It will therefore equal $D_{d0}$ to 12 digits if $p \sim
10^{-12} \, \Lambda$.  This limit also applies to the integrand itself,
so we may replace $f_{h0}$ with $f_{d0}$ for values of $p$ less than
${\cal O}(10^{-12} \, \Lambda)$.  In practice however, we use a limit
of $p = 10^{-17} \, \Lambda$ to ensure that we have enough points below
this region to fit the cosine behavior.

\subsection{New Integration Limits}

For the case of $f_{h0}$, we have limited the range of integration to
be $10^{-17} \, \Lambda$ to $\infty$.  (Below this range, we use
$f_{d0}$.)  Naturally, we cannot integrate to infinity and instead must
find a new limit to replace it.  Let us call this limit $\lambda$.  We
choose $\lambda$ such that
\begin{equation}
\frac{p^2}{2 \pi^2 D_{h0}(p,\Lambda)} \int_{\lambda}^{\infty} dq \, 
  \int_{-1}^{1} dz \, 
  \frac{U_2\left(\vec{q} + \frac{1}{2}\vec{p}\right) U_2\left(\vec{p} 
     + \frac{1}{2}\vec{q}\right)}{2p^2 + 2q^2 + 2pqz} 
     f_{h0}(q, \Lambda) < {\cal O}(10^{-12}).
\end{equation}

The exponentials in $U_2$ suggest that the integrand should die off
quickly, allowing us to make an initial guess for $\lambda$ using
\begin{equation}
\me^{- \lambda^2/\Lambda^2} = 10^{-12} .
\end{equation}
\noindent This gives an initial value of $\lambda \simeq 5.25 \,
\Lambda$.  However, the double integral makes the analysis harder since
we cannot determine the exact behavior.  We must resort to numerical
computation, and some sample calculations reveal that a limit of
$\lambda = 10 \, \Lambda$ is sufficient for our purposes.

From $10^{-17} \, \Lambda$ down to 0, $f_{h0}$ is replaced by
$f_{d0}$.  We would like to replace the $0$ limit with a larger value
that still maintains our desired accuracy.  Even though we know the
analytic solution for $f_{d0}$, narrowing the range of integration will
reduce our computational effort.  Call this new lower limit $\epsilon$,
which is chosen so that
\begin{equation}
\frac{p^2}{2 \pi^2 D_{h0}(p,\Lambda)} \int_{0}^{\epsilon} dq \, 
    \int_{-1}^{1} dz \, \frac{U_2\left(\vec{q} 
    + \frac{1}{2}\vec{p}\right) U_2\left(\vec{p} 
    + \frac{1}{2}\vec{q}\right)}{2p^2 + 2q^2 + 2pqz} 
    f_{d0}(q, \Lambda) < {\cal O}(10^{-12}) .
\end{equation}
\noindent We assume that $\epsilon \ll 10^{-17} \, \Lambda < p$.  To
within 12 digits of accuracy, the $z$ integration can be written as
\begin{eqnarray}
\int_{-1}^{1} dz \, \frac{U_2\left(\vec{q} 
   + \frac{1}{2}\vec{p}\right) U_2\left(\vec{p} 
   + \frac{1}{2}\vec{q}\right)}{2p^2 + 2q^2 + 2pqz} 
   & = & \int_{-1}^{1} dz \, \frac{U_2\left(\vec{p}/2\right) 
   U_2\left(\vec{p}\right)}{2p^2 + 2pqz} \nonumber
\\
& = & \frac{1}{2pq} \, U_2\left(\vec{p}/2\right) 
    U_2\left(\vec{p}\right) \, \ln\left(\frac{p^2 + pq}{p^2 - pq}\right) .
\end{eqnarray}
\noindent Since $q \ll p$, the logarithm can be approximated as
$2q/p$.  Our constraint for $\epsilon$ now becomes
\begin{equation}
\frac{1}{2 \pi^2 D_{h0}(p,\Lambda)} \,  
  U_2\left(\vec{p}/2\right) U_2\left(\vec{p}\right) \, 
  \int_0^{\epsilon} dq \, f_{d0}(q) < {\cal O}(10^{-12}).
\end{equation}
\noindent The values of $p$ are of the same order as $\Lambda$, so we
expect the value of $U_2$ to be ${\cal O}(1)$.  The function $f_{d0}$
is also ${\cal O}(1)$, so it is simply replaced by $1$ in this
approximation.  This leaves an integral with a value of $\epsilon$. 
When combined with $D_{h0}(p,\Lambda) \sim {\cal O}(p)$, we find that
$\epsilon/p < {\cal O}(10^{-12})$.  The smallest value for $p$ is
$10^{-17} \, \Lambda$, implying that $\epsilon \simeq 10^{-29} \,
\Lambda$.  In practice, this value is sufficient for 12 digits of
accuracy.

Having replaced the limits for $f_{h0}$, we move on to $f_{l0}$. 
Again, we must find a finite upper limit to substitute for infinity. 
For $p > 10^{15} \, \eta_3$, $f_{l0}$ is replaced by $f_{d0}$.  This
new limit, $\lambda$, is determined by the condition
\begin{eqnarray}
&& \hspace{-1.5in} \frac{(p^2 + \eta_3^2 - \eta_2^2)}
                        {4 \pi^2 p \, D_{l0}(\eta_2,\eta_3,p)} 
  \int_{\lambda}^{\infty} dq \: \left[ \frac{1}{q} 
  \ln\left(\frac{p^2 + q^2 + pq}{p^2 + q^2 - pq}\right) f_{d0}(q) \right] 
  \nonumber
\\
&& \simeq  \frac{(p^2 + \eta_3^2 - \eta_2^2)}
                {4 \pi^2 p \, D_{l0}(\eta_2,\eta_3,p)} 
   \int_{\lambda}^{\infty} \frac{dq}{q} \: \left(\frac{2p}{q}\right) 
     f_{d0}(q) \nonumber
\\
&& = \frac{(p^2 + \eta_3^2 - \eta_2^2)}{2 \pi^2 \, D_{l0}(\eta_2,\eta_3,p)} 
     \int_{\lambda}^{\infty} \frac{dq}{q^2} \: f_{d0}(q) \nonumber
\\
&& \sim \frac{(p^2 + \eta_3^2 - \eta_2^2)}{2 \pi^2 \, D_{l0}(\eta_2,\eta_3,p)} 
    \int_{\lambda}^{\infty} \frac{dq}{q^2} \nonumber
\\
&& = \frac{(p^2 + \eta_3^2 - \eta_2^2)}{2 \pi^2 \, D_{l0}(\eta_2,\eta_3,p)} \, 
    \left(\frac{1}{\lambda}\right)  \le {\cal O}(10^{-12}) \label{eqn:fl0UL},
\end{eqnarray}
\noindent where we have used the fact that $q \gg p \sim \eta_3$.  For
values of $p$ much larger than $\eta_3$, the term in
Eq.~(\ref{eqn:fl0UL}) is proportional to $p/\lambda$.  The largest
value $p$ can obtain is $10^{15} \, \eta_3$, implying $\lambda =
10^{27} \, \eta_3$.  Numerical calculations verify that this limit is
sufficient to assure the desired accuracy.

Finally, we must replace the lower limit for $f_{l0}$ with a non-zero
value $\epsilon$ that we assume to be much smaller than $\eta_3$.  Our
requirement is that
\begin{equation}
 \frac{(p^2 + \eta_3^2 - \eta_2^2)}{4 \pi^2 p 
    (\eta_3^2 - \eta_2^2) D_{l0}(\eta_2,\eta_3,p)} \int_0^{\epsilon} dq \, 
    q \ln\left(\frac{\eta_3^2 + 2p^2 + 2pq}{\eta_3^2 + 2p^2 - 2pq}\right) 
    f_{l0}(\eta_2,\eta_3,q) < {\cal O}(10^{-12}) .
\end{equation}
\noindent Expanding the logarithm to ${\cal O}(q)$ yields
\begin{equation}
 \frac{(p^2 + \eta_3^2 - \eta_2^2)}{\pi^2 (\eta_3^2 - \eta_2^2) 
     (\eta_3^2 + 2 p^2) D_{l0}(\eta_2,\eta_3,p)} 
     \int_0^{\epsilon} dq \, q^2  \, f_{l0}(\eta_2,\eta_3,q) 
     < {\cal O}(10^{-12}) \label{eqn:fl0LL}.
\end{equation}
\noindent For small values of $q$, we will find that $f_{l0}$ is
approximately constant and of ${\cal O}(1)$.  Therefore, the integral
is roughly equal to $\epsilon^3$.  If $p \ll \eta_3$ or $p \sim
\eta_3$, then Eq.~(\ref{eqn:fl0LL}) is ${\cal O}(\epsilon^3/\eta^3)$. 
If $p \gg \eta_3$, then it is ${\cal O}(\epsilon^3/p\eta_3^3) < {\cal
O}(\epsilon^3/\eta^3)$.  This seems to imply that a value of $\epsilon
\simeq 10^{-4} \, \eta_3$ is adequate.

Unfortunately, using this value of $\epsilon$ will result in poor
accuracy when $\eta_2 \simeq \eta_3$.  This is a result of the
$(\eta_3^2 - \eta_2^2)$ term in the denominator of
Eq.~(\ref{eqn:fl0LL}).  Originally, this was our approximation to the
term $(q^2 + \eta_3^2 - \eta_2^2)$ in the integral equation.  When the
energies are nearly equal, our approximation needs to be $q^2$.  The
condition on $\epsilon$ should now become
\begin{equation}
 \frac{(p^2 + \eta_3^2 - \eta_2^2)}{\pi^2 (\eta_3^2 + 2 p^2) D_{l0}
 (\eta_2,\eta_3,p)} \int_0^{\epsilon} dq   
 \, f_{l0}(\eta_2,\eta_3,q) < {\cal O}(10^{-12}) \label{eqn:newfl0LL}.
\end{equation}
\noindent The integral is roughly equal to $\epsilon$, and the entire
term is ${\cal O}(\epsilon/\eta)$. This means we must use the lower
limit $\epsilon = 10^{-12} \, \eta_3$.

These derivations are general enough to determine the appropriate
limits for other cases of desired accuracy.  Keep in mind however that
the limits must always be tested numerically to ensure that they are
indeed sufficient.

\subsection{Discrete Point Spacing}

Now that we have limits in place, we must choose the discrete values of
$q$ within these limits at which to evaluate our functions.  We employ
discrete points that are equally spaced on a logarithmic scale.  There
are two main reasons for making this choice.

First, we have already seen that $f_{d0} = \cos\left(s_0 \ln(p/\Lambda)
+ \theta \right)$ is periodic on a logarithmic scale.  In fact, other
functions have similar behavior, including $G_3$.  It makes sense that
a logarithmic spacing is suited to capturing the behavior of the
system.

\begin{figure}
\begin{center}
\includegraphics[width=5.0in]{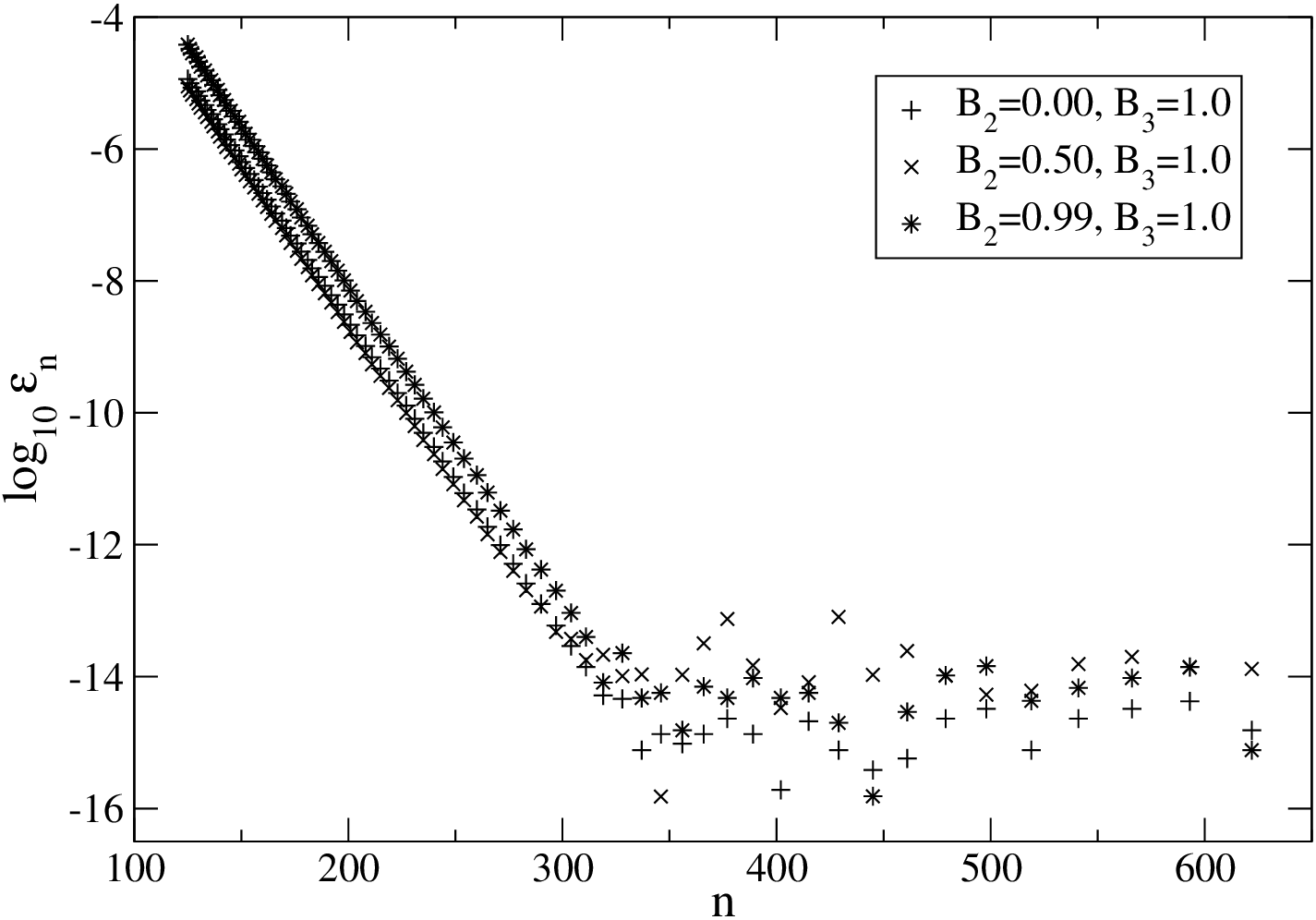}
\end{center}
\vspace*{-.1in}
\caption{\label{fig:PhaseError} Relative error in the phase 
      appearing in pseudo-wave functions as a function of the number 
      of points used in the discretized approximation for several pairs 
      of two-body and three-body binding energies.}
\end{figure}

Second, and far more important, by choosing points in this manner an
integration of $q$ from $0$ to $\infty$ becomes an integration of
$\ln(q)$ from $-\infty$ to $\infty$.  By equally spacing points on this
log scale, we can achieve convergence that improves {\it exponentially}
with the spacing \cite{Tadmor}, as opposed to the more typical power law
convergence.  This drastically reduces the number of points needed to
achieve our desired accuracy.  For instance, to cover the range of
$f_{l0}$ from $10^{-12} \, \eta_3$ to $10^{15} \, \eta_3$, and maintain
12 digits of accuracy, we can space our points by $p_{n+1} = p_n
\me^{0.2}$.  This requires only about 300 points. Exponential
convergence with the number of point is extremely powerful and it is
not widely appreciated that this is ever possible. A simple
demonstration (with minor errors) can be found in an appendix in Mohr's
thesis \cite{Mohr03}. Figure \ref{fig:PhaseError} illustrates this
exponential convergence which eventually terminates due to roundoff
errors.

\section{Discretized Equations}

The discretized integral equations are shown below.  Here, $\Delta$
represents the momentum spacing, and the momentum values are related by
the equation $p_{n+1} = p_{n} \, \me^{\Delta}$.  The identity matrix is
represented by $I_{nm}$, and the ceiling function represented by
$\lceil x \rceil$ returns an integer value $n$ such that $(n-1) < x \le
n$.

{\bf Low-Momentum:} These are easily derived from Eq.~(83). The value
of $n$ for $p_n$ ranges from 0 to $N_{mid}$, while the value of $m$ for
$q_m$ ranges from 0 to $N_{max}$. These are defined as: $p_0 = 10^{-12}
\, \eta_3$,  $N_{mid} = \lceil \ln(10^{27})/\Delta \rceil$, and
$N_{max} = \lceil \ln(10^{39})/\Delta \rceil$ .

\begin{equation}
 \sum_{m = 0}^{N_{mid}} \left( M_{nm} - I_{nm} \right) \, f_{l0}(p_m) = b_n ,
\end{equation}
\begin{equation}
M_{nm} = \frac{\Delta}{4 \pi^2 D_{l0}(\eta_2,\eta_3,p_n)} 
   \frac{q_m^2 (p_n^2 + \eta_3^2 - \eta_2^2)}
        {p_n (q_m^2 + \eta_3^2 - \eta_2^2)} 
        \ln\left(\frac{\eta_3^2 + 2 p_n^2 + 2 q_m^2 + 2 p_n q_m}
                      {\eta_3^2 + 2 p_n^2 + 2 q_m^2 - 2 p_n q_m}\right) ,
\end{equation}
\begin{eqnarray}
&& \hspace{-0.5in} b_n = - 
    \frac{\Delta (p_n^2 + \eta_3^2 - \eta_2^2)}
         {4 \pi^2 p_n D_{l0}(\eta_2,\eta_3,p_n)} 
    \sum_{m = N_{mid}+1}^{N_{max}} 
      \left[ \ln\left(\frac{\eta_3^2 + 2 p_n^2 + 2 q_m^2 + 2 p_n q_m}
                           {\eta_3^2 + 2 p_n^2 + 2 q_m^2 - 2 p_n q_m}\right) 
       \right. \nonumber
\\
&& \hspace{2in} \left. \times 
  \cos\left( s_0 \ln\left(\frac{q_m}{\Lambda}\right) + \theta \right) \right].
\end{eqnarray}

{\bf High-Momentum:} The value of $n$ for $p_n$ ranges from 0 to
$N_{max}$, while the value of $m$ for $q_m$ ranges from $-N_{min}$ to
$N_{max}$.  These are defined as: $p_0 = 10^{-17} \, \Lambda$, 
$N_{max} = \lceil \ln(10^{18})/\Delta \rceil$, and $N_{min} = \lceil
\ln(10^{12})/\Delta \rceil$ .

\begin{equation}
 \sum_{m = 0}^{N_{max}} \left( M_{nm} - I_{nm} \right) \, f_{h0}(p_m) = a_n 
    + \delta_0 \, b_n,
\end{equation}
\begin{equation}
M_{nm} = \frac{q_m p_n^2 \Delta}{2 \pi^2 D_{h0}(p_n,\Lambda)} K(p_n, q_m),
\end{equation}
\begin{equation}
a_n = - \frac{p_n^2 \Delta}{2 \pi^2 D_{h0}(p_n,\Lambda)} 
   \sum_{m = -N_{min}}^{-1} q_m \, K(p_n,q_m) 
   \cos\left( s_0 \ln\left(\frac{q_m}{\Lambda}\right) + \theta \right),
\end{equation}
\begin{equation}
b_n = \frac{p_n^2 D_{1h0}(p_n,\Lambda)}{\Lambda^2 G_{2,0} D_{h0}(p_n,\Lambda)},
\end{equation}
\begin{eqnarray}
K(p_n, q_m) & = & \frac{\Delta}{2 \Lambda^2} \sum_{k = -200}^{200} 
   \left[ (1 - z_k^2) \left( 1 + h_2 \left( \frac{q_m^2}{\Lambda^2} 
   + \frac{q_m p_n z_k}{\Lambda^2} + \frac{p_n^2}{4 \Lambda^2} \right)\right) 
   \right. \nonumber
\\
&& \times \left( 1 + h_2 \left( \frac{p_n^2}{\Lambda^2} 
   + \frac{q_m p_n z_k}{\Lambda^2} + \frac{q_m^2}{4 \Lambda^2} \right)\right), 
   \nonumber
\\
&& \left. \times 
  \frac{\exp\left( -(5 p_n^2 + 8 p_n q_m z_k + 5 q_m^2)/(4 \Lambda^2) \right)}
       {2\left( p_n^2/\Lambda^2 + q_m^2/\Lambda^2 + p_n q_m z_k/\Lambda^2 
          \right)} \right],
\end{eqnarray}
\begin{equation}
z_k = \frac{\me^{0.2 \, k} - 1}{\me^{0.2 \, k} + 1}.
\end{equation}
The hyperbolic tangent discretization we use for the angular integral,
with 401 angular points, has been numerically verified to exceed our
needs over the entire range of binding energies we explore. There may
be equally good or better choices.


\section{Analytic and Numerical Results}
\label{results}

A computer program to solve the discretized integral equations can be
implemented with the help of some basic numerical algorithms.  Once
completed, it can be used to generate highly accurate results and
analyze the three-body system.  Nonetheless, attempts to extract
analytic results should not be overlooked.  Even some of the most
general properties of the integral equations allow us to draw
conclusions about the behavior of the system.

We begin this chapter with analytic results obtained from studying the
leading order integral equations.  These results include statements
about the cutoff dependence of bound-state energies and the phase for
$f_{d0}$.  A proof for the cyclic behavior of $\delta_0$ is given,
which is then used to infer similar behavior for $G_3$.

Following the analytic results is a section containing numerical
solutions to the integral equations.  Here we examine behavior that
cannot be determined analytically.  Solutions for the functions
$f_{l0}$ and $f_{h0}$ are shown, and the cutoff dependence of $G_3$ is
calculated.  Some relations that are proven analytically are also
verified numerically.



In Section \ref{sec:sol_inteq}, we saw that Eq.~(\ref{eqn:QRfl0})
contains no $\Lambda$ dependence, making the function $f_{l0}(p)$
cutoff-independent.  However, for values of $p \gg \eta_3$, the
solution must behave like $f_{d0}(p) = \cos\left(s_0 \,
\ln\left(p/\Lambda\right) + \theta \right)$, which explicitly has the
cutoff in it.  The only way these two equations can be reconciled is if
$\theta$ contains cutoff dependence of the form $s_0 \, \ln(\Lambda)$. 
Any remaining part of $\theta$ must be a function of $\eta_2$ and
$\eta_3$.  Therefore, we can write
\begin{equation}
\theta = s_0 \, \ln\left(\Lambda/\eta_3\right) 
  + \tilde{\theta}\left(\eta_2/\eta_3\right) \label{eqn:theta_tilde},
\end{equation}
\noindent where $\tilde{\theta}$ is a dimensionless function of the
ratio $\eta_2/\eta_3$.  This relation holds for any values of $\eta_2$
and $\eta_3$, including all $\eta_3$ values corresponding to multiple
bound states with the same $\eta_2$.  Using $\eta_3$ in the ratio with
$\Lambda$ is a matter of choice. Any quantity composed of $\eta_2$ and
$\eta_3$ with the same dimension as $\Lambda$ would work just as well;
however, we need to allow $\eta_2 = 0$, so $\eta_2$ alone is a poor
choice.

While Eq.~(\ref{eqn:theta_tilde}) is quite simple, it has many
interesting consequences.  As we mentioned earlier, two different
three-body bound states in the same spectrum must have the same phase
$\theta$.  Suppose that we choose some fixed values for $\eta_2$ and
$\eta_3$ and make them cutoff-independent by choosing the appropriate
$\Lambda$ dependence for $G_2$ and $G_3$.  This might be desirable if
we are trying to match those energies to experimental data.  The phase
for the solution in this case is
\begin{equation}
\theta_{\Lambda} =  s_0 \, \ln\left(\Lambda/\eta_3\right) 
   + \tilde{\theta}\left(\eta_2/\eta_3\right) ,
\end{equation}
\noindent for some given cutoff $\Lambda$.  If the cutoff is changed,
$G_2$, $G_3$, and $\theta_{\Lambda}$ will all change with it, but
$\eta_2$, $\eta_3$, and $\tilde{\theta}(\eta_2/\eta_3)$ will not.

Imagine now that we find a second three-body bound-state solution with
the same phase.  Let us call its energy $\bar{\eta}_3$.  The phase for
this solution is
\begin{equation}
\bar{\theta}_{\Lambda} =  s_0 \, \ln\left(\Lambda/\bar{\eta}_3\right) 
     + \tilde{\theta}\left(\eta_2/\bar{\eta}_3\right) ,
\end{equation}
\noindent which must be equal to $\theta_{\Lambda}$ by assumption. 
This results in the relation
\begin{equation}
s_0 \, \ln\left(\bar{\eta}_3/\eta_3\right) = 
   \tilde{\theta}\left(\eta_2/\bar{\eta}_3\right) 
   - \tilde{\theta}\left(\eta_2/\eta_3\right) .
\end{equation}
\noindent If the cutoff is now changed to a new value $\Lambda'$, the
original data gives a phase of
\begin{equation}
\theta_{\Lambda'} =  s_0 \, \ln\left(\Lambda'/\eta_3\right) 
    + \tilde{\theta}\left(\eta_2/\eta_3\right) .
\end{equation}
\noindent The question is whether $\bar{\eta}_3$ is still a valid
solution.  Its new phase is
\begin{eqnarray}
\bar{\theta}_{\Lambda'} & = & s_0 \, \ln\left(\Lambda'/\bar{\eta}_3\right) 
   + \tilde{\theta}\left(\eta_2/\bar{\eta}_3\right) \nonumber
\\
& = & s_0 \, \ln\left(\Lambda'/\bar{\eta}_3\right) 
   + \left[ s_0 \, \ln\left(\bar{\eta}_3/\eta_3\right) 
   + \tilde{\theta}\left(\eta_2/\eta_3\right) \right] \nonumber
\\
& = & s_0 \, \ln\left(\Lambda'/\eta_3\right) 
   + \tilde{\theta}\left(\eta_2/\eta_3\right) \nonumber
\\
& = & \theta_{\Lambda'} .
\end{eqnarray}
\noindent Since the phases still match, $\bar{\eta}_3$ is still a
bound-state solution.  This remains true for any cutoff, implying that
$\bar{\eta}_3$ is also cutoff-independent like $\eta_3$.  Of course,
the same statement applies to any other three-body bound state {\em
making the entire spectrum completely independent of $\Lambda$}.  Such
behavior should come as no surprise since the leading order equations
represent the $\Lambda \rightarrow \infty$ limit.  Keep in mind that
this is true for any other physical quantity, but does not apply to the
couplings.  Obviously, it is the cutoff dependence of the couplings
that enables the bound states to be cutoff independent.  We have shown
that a single three-body contact interaction allows us to renormalize
the entire three-body bound-state spectrum.

Just as $f_{l0}$ has no dependence on $\Lambda$, neither does it have
any dependence on $h_2$.  As a consequence, $h_2$ has no effect at
leading order on the binding energies or other physical quantities.  It
does appear in the equation for $f_{l1}$ however, showing that it is
needed when considering first order corrections.  Because we work only
to leading order, we use $h_2 = 0$ below unless stated otherwise.

We now turn to the equation for $f_{h0}(p)$:
\begin{eqnarray}
f_{h0}(p, \Lambda) = \frac{p^2}{2 \pi^2 D_{h0}(p,\Lambda)} 
  \int_0^{\infty} dq \, \int_{-1}^{1} dz \, 
  \frac{U_2\left(\vec{q} 
    + \frac{1}{2}\vec{p}\right) U_2\left(\vec{p} 
    + \frac{1}{2}\vec{q}\right)}
       {2p^2 + 2q^2 + 2pqz} f_{h0}(q, \Lambda) \nonumber
\\
 \hspace{0.5in} -~\delta_0 \, \frac{3 h_2^2 + 8 h_2 + 16}
       {128 \sqrt{2} \, \pi^{3/2}} \left( \frac{D_{1h0}(p,\Lambda)}
       {D_{h0}(p,\Lambda)} \right) \left( \frac{p^2}{\Lambda^2} \right) 
       \label{eqn:thisfh0}.
\end{eqnarray}
\noindent The dependence on $p$ and $\Lambda$ has been explicitly
shown.  Since $f_{h0}$ is dimensionless, only the ratio $p/\Lambda$ can
occur in the function.  Furthermore, there is a dependence upon $h_2$
and $\theta$.  The $h_2$ dependence comes from its appearance in the
function $U_2(p)$, either directly in the integral or indirectly in
$D_{h0}$ and $D_{1h0}$.  The $\theta$ dependence is a result of
$f_{h0}$ approaching $f_{d0}$ for $p \ll \Lambda$.

Assuming that $\Lambda$ is held fixed, choosing a value for $\delta_0$
will uniquely determine the phase.  Thus, we can view the phase as a
function of the coupling, $\theta(\delta_0)$.  Conversely, choosing a
phase determines the coupling, so it is equally valid to treat the
coupling as a function of the phase, $\delta_0(\theta)$.  The coupling
$\delta_0$ can also have a dependence upon $h_2$ but not upon
$\Lambda$.  The reason is that $\delta_0$ is dimensionless and there is
no other quantity available to form a dimensionless ratio with
$\Lambda$.

With this in mind, consider a solution to Eq.~(\ref{eqn:thisfh0}) with
a coupling of $\delta_0(\theta)$ and a function $f_{h0}(p)$ that
behaves asymptotically as $\cos\left(s_0 \, \ln\left(p/\Lambda\right) +
\theta \right)$ for small $p$.  For a phase of $\theta + 2 \pi$, the
cosine behavior will remain unchanged.  We may therefore conclude that 
\begin{equation}
\delta_0(\theta + 2 \pi) = \delta_0(\theta) ,
\end{equation}
\noindent showing that the coupling exhibits periodic behavior.  If we
apply the operator $-\partial^2/\partial\theta^2$ to both sides of
(\ref{eqn:thisfh0}), we find once again that the asymptotic cosine
behavior is the same.  This implies
\begin{equation}
-\frac{\partial^2 \delta_0}{\partial\theta^2} = \delta_0(\theta) .
\end{equation}
\noindent The conclusion to be drawn is that the coupling may be
written as a cosine function with some amplitude ${\cal A}$ and phase
$\phi$.  These two parameters will contain any $h_2$ dependence that
$\theta$ may possess, so we shall write
\begin{equation}
\delta_0(\theta) = {\cal A}(h_2) \, 
   \cos\left( \theta + \phi(h_2) \right) \label{eqn:PseudoCosine}.
\end{equation}

The equation relating $\delta_0$ and $G_3$,
\begin{equation}
G_{3,0}  =  \frac{\delta_0}{G_{2,0} \, {\cal I}_0 - D_{2,0}\,
    \delta_0/\Lambda^4 } ,
\end{equation}
\noindent shows that whenever $\delta_0$ is zero, so is
$G_3$.\footnote{The only problem that might arise is if ${\cal I}_0 =
0$ in such a way that the ratio is non-zero, but we have found no
parameters for which this is true.}  Since two adjacent zeros of
$\delta_0$ occur when the cutoffs satisfy
\begin{equation}
\theta_{\Lambda'} - \theta_{\Lambda} = s_0 \, 
  \ln\left(\frac{\Lambda'}{\Lambda}\right) = \pi ,
\end{equation}
\noindent the adjacent zeros of $G_3(\Lambda)$ should be spaced by a
cutoff factor of $\Lambda'/\Lambda = \exp(\pi/s_0) \simeq
22.69438259536$.  This suggests that $G_3(\Lambda)$ may also possess
cyclic behavior, but this must be verified numerically, which is done
below.



We begin our numerical investigation by considering solutions for the
functions $f_{l0}$ and $f_{h0}$ which lead to an approximation for the
complete function $f(p)$.  Next, the cutoff independence of the
three-body spectrum is verified, followed by an analysis of the
coupling constants.  These numerical results are confirmed by Wilson's
calculations \cite{Wil00}.

\subsection{Pseudo-Wave Functions}

\begin{figure}
\begin{center}
\includegraphics[width=5.0in]{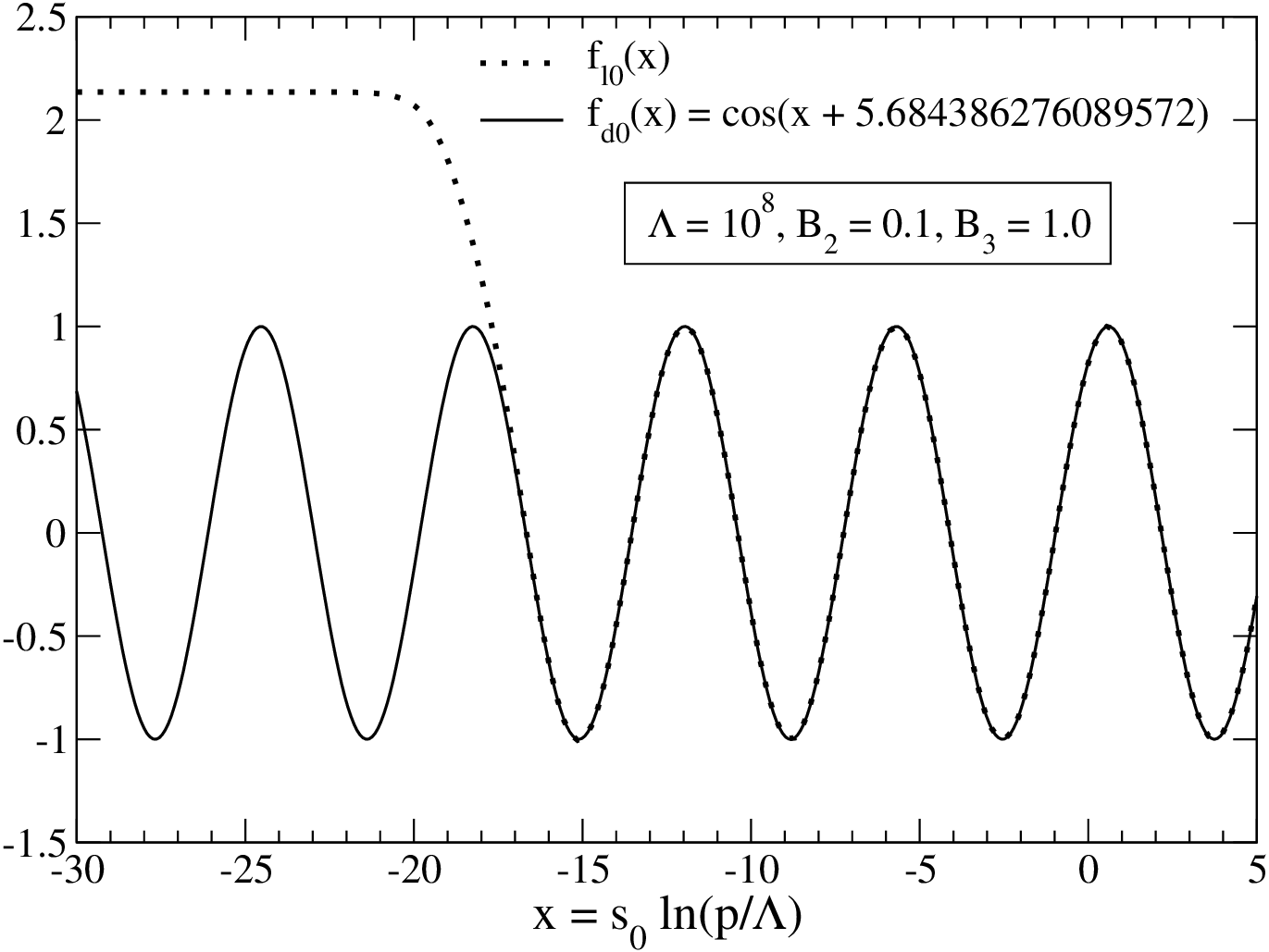}
\end{center}
\vspace*{-.1in}
\caption{Numerical solution of $f_{l0}(p)$ shown
matching on to $f_{d0}(p)$ for the case of $B_2 = 0.1$, $B_3 = 1.0$,
and $\Lambda = 10^8$.}
\label{fig:FL0-2} 
\end{figure}

Figure \ref{fig:FL0-2} shows the numerical solution for $f_{l0}(p)$ in
the case of $B_2 = 0.1$, $B_3 = 1.0$, and $\Lambda = 10^8$.  Notice
that it is constant for small values of momentum and then takes on the
cosine behavior of $f_{d0}$ as $p$ becomes large.  Since the behavior
of the solution to the low-momentum equation is determined by the ratio
$B_2/B_3$, this figure is representative of all solutions, as
demonstrated explicitly in Mohr's thesis \cite{Mohr03}. 

\begin{figure}[p]
\begin{center}
\includegraphics[width=5.0in]{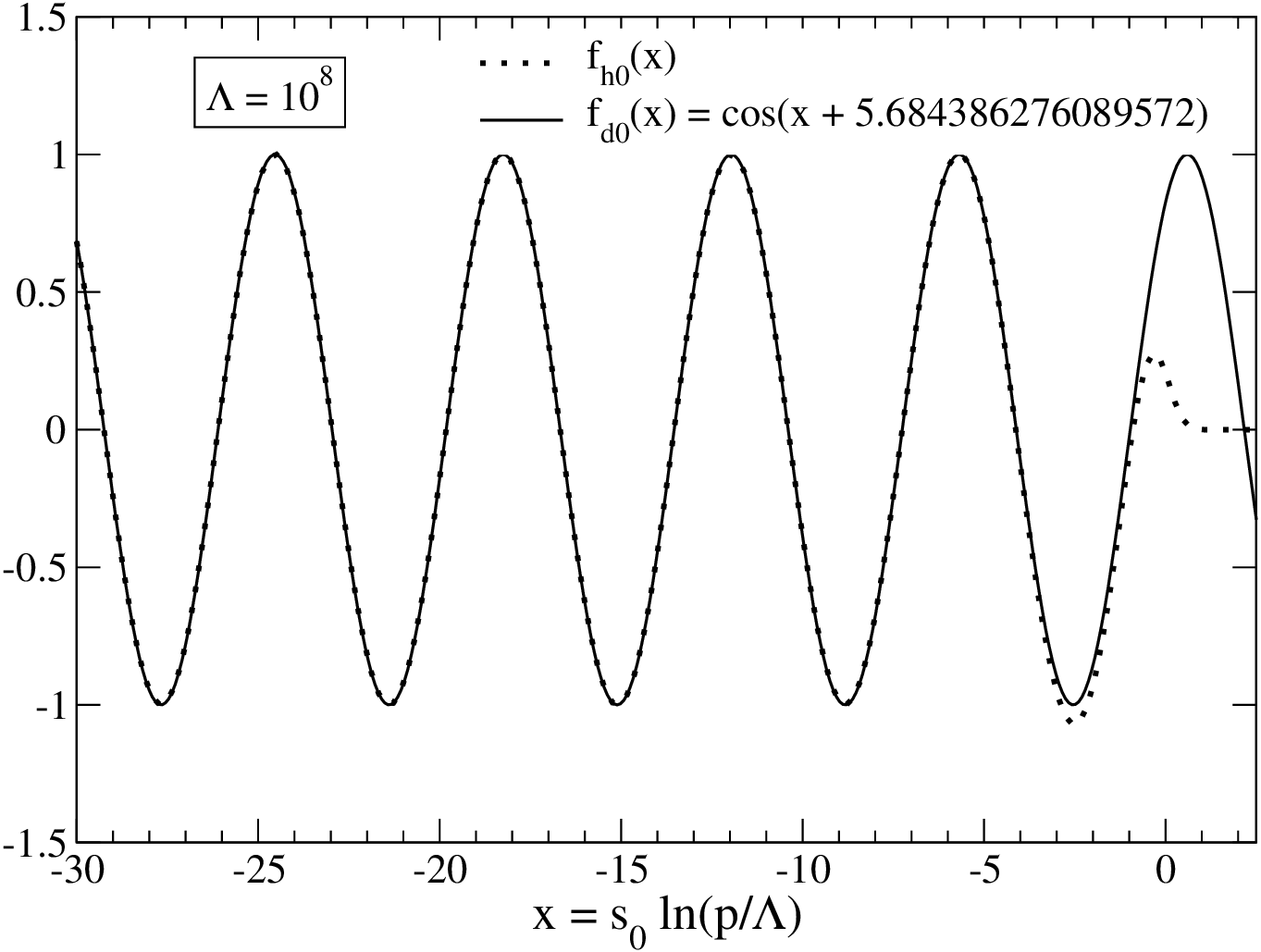}
\end{center}
\vspace*{-.1in}
\caption{Numerical solution of $f_{h0}(p)$ for the
case $\theta = 5.684386276089572$ and $\Lambda = 10^8$.  The dashed
line is the best-fit cosine curve that matches the low-momentum
behavior.}
\label{fig:FH0-2} 
\vspace*{.5in}
\begin{center}
\includegraphics[width=5.0in]{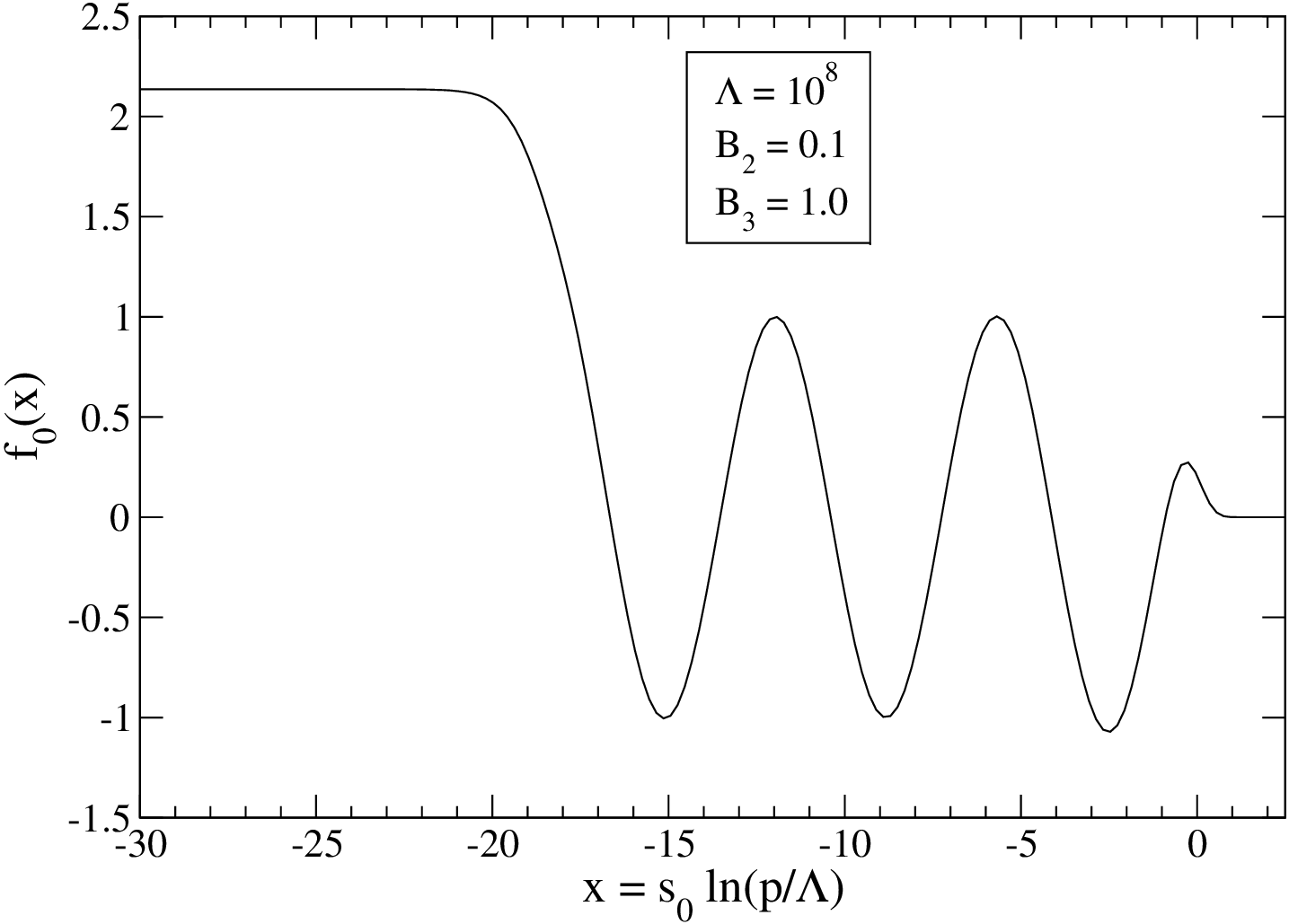}
\end{center}
\vspace*{-.1in}
\caption{Numerical solution of $f_0(p)$ for the case
of $B_2 = 0.1$, $B_3 = 1.0$, and $\Lambda = 10^8$.}
\label{fig:F0-2} 
\end{figure}

This binding energy leads to $\theta = 5.684386276089572...$, and with
$\Lambda = 10^8$ we find the numerical solution for the function
$f_{h0}(p)$ shown in Figure~\ref{fig:FH0-2}.  This is the high-momentum
function corresponding to the low-momentum function in
Figure~\ref{fig:FL0-2}.  For all $f_{h0}$ solutions, we see the cosine
behavior for $p \ll \Lambda$ and a suppression of large momentum values
when $p > \Lambda$.  This suppression is an effect of the Gaussian
behavior of $U_2$.

Combining $f_{l0}$, $f_{d0}$ and $f_{h0}$, we see the overall leading
order behavior for the function $f(p)$.  Figure~\ref{fig:F0-2} shows
$f_{l0}+ f_{h0} - f_{d0}$ for $B_2 = 0.1$, $B_3 = 1.0$, and $\Lambda =
10^8$.  All solutions show this same qualitative behavior, with a flat
plateau at low momenta going to cosine behavior for medium momenta and
finally decaying rapidly at large momenta above the cutoff. Increasing
the cutoff while changing the coupling so that the phase remains
constant simply extends the cosine region, pushing the eventual decay
to higher momenta.

\begin{figure}
\begin{center}
\includegraphics[width=5.0in]{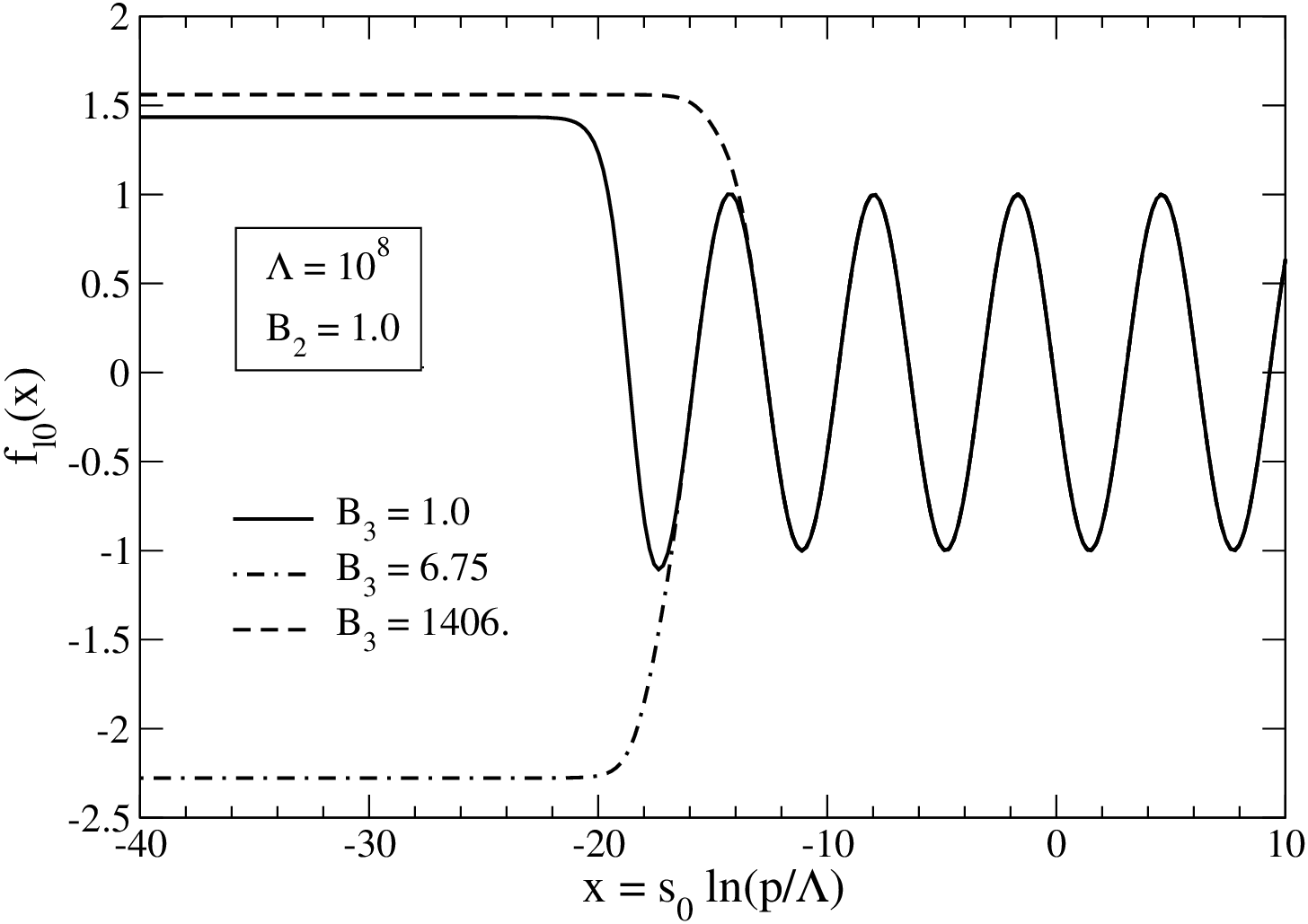}
\end{center}
\vspace*{-.1in}
\caption{Numerical solution of $f_{l0}(p)$
for three bound states using $B_2 = 1.0$ and $\Lambda = 10^8$. The
three-body bound state energies are 1.0, 6.7502901502599 and
1406.1303932044.}
\label{fig:FL0-Spectrum2} 
\end{figure}

Several bound states typically exist for the same values of $B_2$ and
$\Lambda$.  In the case of $B_2 = 1.0$ and $B_3 = 1.0$ for $\Lambda =
10^8$, there exist states of energy $B_3 = 6.7502901502599$ and $B_3 =
1406.130393204$.  The solutions of $f_{l0}$ for these energies are
shown in Figure~\ref{fig:FL0-Spectrum2}.  All of the functions are very
similar.  The only real difference is in the length of the initial
plateau.  As the bound-state energies become larger, the flat region
becomes longer, joining the cosine near later peaks. For a fixed cutoff
there are a finite number of solutions because the cosine terminates
near the cutoff, but as the cutoff is extended and further periods of
the cosine result, additional bound states appear.

\subsection{Three-Body Binding Energies}

Above we proved that the three-body bound-state spectrum is cutoff
independent to leading order.  We choose $B_2 = 0$ (for which there is
a limit cycle that persists to arbitrarily small cutoffs) and let $G_3$
change with $\Lambda$ so that the state $B_3 = 1.0$ is held constant. 
For this special case there are an infinite number of low-lying bound
states as shown by Efimov \cite{Efi70,Efi71}. Two other states, one
shallower and one deeper, are calculated as the cutoff changes.  Since
very small fluctuations are impossible to see in a plot, Table
\ref{tab:B3vsLAMBDA-1} shows the calculated energies for several values
of $\Lambda$.  Using Efimov's result that the ratio of adjacent binding
energies is $\exp(2 \pi/s_0)$ when $B_2 = 0$ \cite{Efi70,Efi71}, the
relative error for each calculation can be determined and is also given
in the table.  This illustrates that each energy is cutoff-independent
to about 12 digits and also matches the true value to the same
accuracy.  As an additional example, Table \ref{tab:B3vsLAMBDA-2} shows
the case of $B_2 = B_3 = 1.0$ and considers the next two deeper states
as $\Lambda$ changes.  The binding energies are approximately 6.75029
and 1406.13.  These energies have been previously calculated by Bedaque
{\em et al.}  \cite{BHK99b} and by Braaten, Hammer and Kusunoki
\cite{BHK03} using a method that gives at most two digits of numerical
precision.  Their results are 6.8 and $1.4\times 10^3$, which match the
results given here to within their relative errors of ${\cal
O}(10^{-3})$. 

\begin{table}
\begin{center}
\begin{tabular}{|c|c|c|c|c|}
\hline
$\Lambda$ & $B_3$ (Shallow) & Error & $B_3$ (Deep) & Error \\
\hline
\hline
100000.00000000 & 0.0019416156131338 & 6.7e-13 & 515.03500138461 & 5.2e-13 \\
738905.60989306 & 0.0019416156131358 & 3.2e-13 & 515.03500138403 & 1.6e-12 \\
5459815.0033144 & 0.0019416156131435 & 4.3e-12 & 515.03500138287 & 3.9e-12 \\
109663315.84284 & 0.0019416156131358 & 3.2e-13 & 515.03500138520 & 6.0e-13 \\
3631550267.4246 & 0.0019416156131435 & 4.3e-12 & 515.03500138520 & 6.0e-13 \\
\hline
\end{tabular}
\end{center}
\caption{Binding energies  and relative errors
for the next shallowest and next deepest 3-body bound states for $B_2 =
0.0$, $B_3 = 1.0$ and various cutoffs. For this case we know the spectrum analytically
and the determination of errors is straightforward.} 
\label{tab:B3vsLAMBDA-1}
\end{table}

\begin{table}
\begin{center}
\begin{tabular}{|c|c|c|}
\hline
$\Lambda$ & $B_3$ \#1 & $B_3$ \#2 \\
\hline
\hline
100000.00000000 & 6.750290150257678 & 1406.13039320296 \\
738905.60989306 & 6.750290150257678 & 1406.13039320593 \\
2008553.6923187 & 6.750290150255419 & 1406.13039320345 \\
14841315.910257 & 6.750290150268966 & 1406.13039320345 \\
298095798.70417 & 6.750290150257678 & 1406.13039320593 \\
5987414171.5197 & 6.750290150259935 & 1406.13039320345 \\
44241339200.892 & 6.750290150257678 & 1406.13039320296 \\
\hline
\end{tabular}
\end{center}
\caption{\label{tab:B3vsLAMBDA-2}Binding energies of the two next
deeper 3-body bound states for $B_2 = B_3 = 1.0$ and various cutoffs.}
\end{table}

\subsection{Couplings}

Equation (\ref{eqn:PseudoCosine}) shows that the coupling $\delta_0$
should have a cosine dependence on the phase $\theta$, which is defined
by $f_{d0} = \cos(s_0\ln(p/\Lambda) + \theta)$.  Numerical data for
$\delta_0$ as a function of the phase show this behavior precisely
\cite{Mohr03}, independent of $B_2$ and $B_3$.

\begin{figure}
\begin{center}
\includegraphics[width=5.0in]{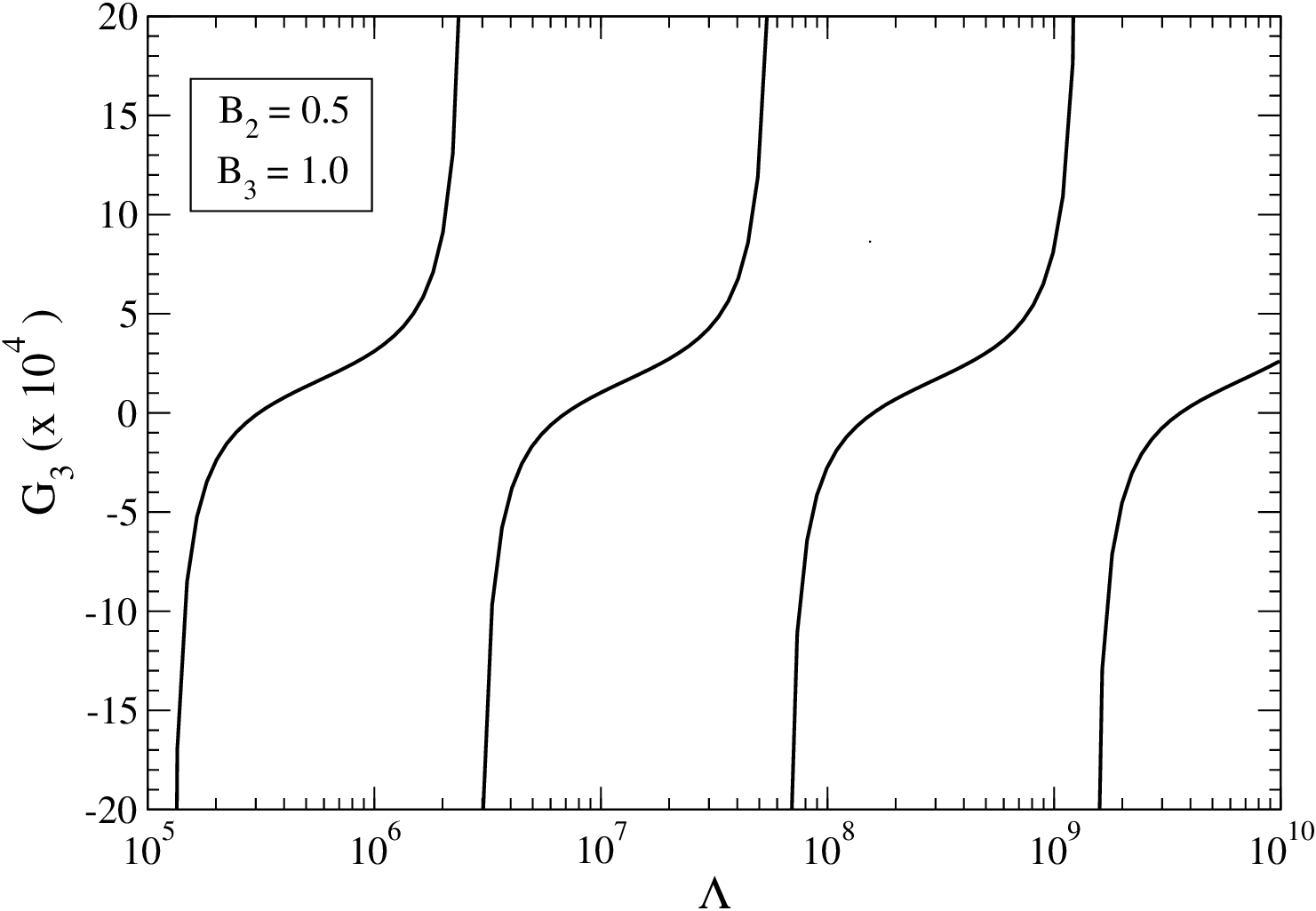}
\end{center}
\vspace*{-.1in}
\caption{The dimensionless three-body coupling
$G_3$ as a function of the cutoff $\Lambda$ for $B_2 = 0.5$ and $B_3 =
1.0$.  The limit-cycle behavior of $G_3$ is evident from the fact that
it is periodic in $\ln(\Lambda)$.}
\label{fig:G3vsLAMBDA-2}
\end{figure}

We suggested above that this periodic behavior should carry over to
$G_3$.  Figure \ref{fig:G3vsLAMBDA-2} displays a plot of $G_3$ as a
function of $\Lambda$.  This data exhibits the limit-cycle behavior of
the three-body coupling.  As the cutoff increases, $G_3$ becomes larger
and larger, eventually diverging to $+ \infty$.  It then jumps to
$-\infty$ and  increases again.  This limit-cycle behavior is not
dependent upon any specific bound-state values, but the positioning of
the cycle is dependent upon the energies. This cyclic behavior has been
previously observed \cite{BHK99,BHK99b} using a sharp cutoff that
simply discards all momenta above $\Lambda$ and a different method for
including the two-body interaction.

\begin{figure}
\begin{center}
\includegraphics[width=5.0in]{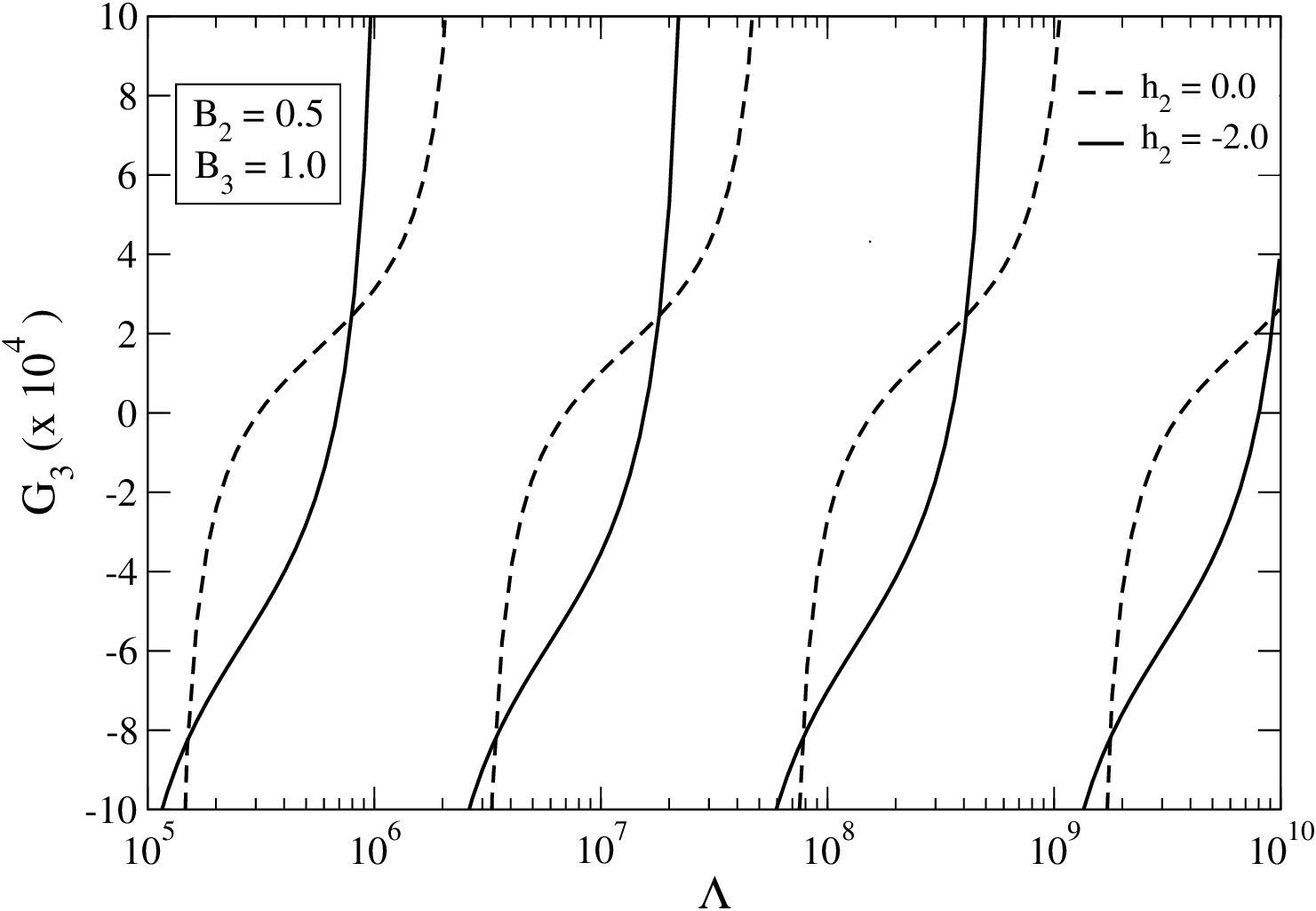}
\end{center}
\vspace*{-.1in}
\caption{\label{fig:G3vsLAMBDA-245}The dimensionless three-body
coupling $G_3$ as a function of the cutoff $\Lambda$ for $B_2 = 0.5$
and $B_3 = 1.0$ with two different values of $h_2$.}
\end{figure}

It is shown above that $h_2$ has no effect on the binding energies to
leading order, but it does affect  $G_3$.  In fact,
Eq.~(\ref{eqn:PseudoCosine}) explicitly exhibits such dependence.  In
Figure~\ref{fig:G3vsLAMBDA-245}  we illustrate the effect on $G_3$ of
changing $h_2$.  The limit cycle behavior persists and the curves are
simply distorted by the presence of non-zero $h_2$. Again, this
behavior has been studied for many cases \cite{Mohr03}.

%


\section{Conclusions}
\label{conclus}

We reduce the equation for S-wave bound states of three bosons
interacting via attractive short range two-body interactions to a set
of coupled integral equations and develop numerical methods that enable
us to solve for both eigenvalues and pseudo-wave functions with a
precision of 11-12 digits. This problem must be regulated (a Gaussian
cutoff here to achieve high numerical precision) and renormalization
produces a well-defined infinite cutoff limit ({\it i.e.}, a continuum
limit) if a three-body contact term is introduced with its coupling
precisely tuned to a limit cycle.

The method of uniformly valid expansion is developed to separate widely
different regions of momenta as a first step towards isolating and then
controlling the high momentum region. We believe that this method will
have applications in other renormalization problems since it addresses
a generic need, but our focus is the renormalization of the three-body
problem.

Unusually high precision is required because this renormalization
problem is intrinsically nonperturbative and must be solved numerically
at this time. We must isolate effects that scale as mixed powers of
${\rm ln}(\Lambda)$ and $1/\Lambda^2$ and this places stringent
requirements on numerical renormalization if one needs to accurately
resolve even a few levels of detail. Irrelevant operators should enable
us to control these sub-leading corrections with a finite number of
couplings, if we can identify appropriate irrelevant operators. It is
simplest to assume that we can employ irrelevant operators from free
field theory ({\it i.e.}, powers of derivatives acting on regulated
delta-functions for the few-body problem) but this needs to be
demonstrated numerically. 

Renormalization replaces an infinite geometric tower of bound states
with effective interactions as the cutoff decreases. Irrelevant
operators must allow us to maintain the least bound states of this
tower as the cutoff descends. Our leading-order calculations  show how
high-momentum effects are funneled through an intermediate-momentum
shell to produce a cutoff-independent tower of bound states that is
entirely determined (with the two-body interaction fixed) by a single
three-body coupling. The uniformly valid expansion provides a tool that
should expose this structure at arbitrarily high levels of precision
while providing direct insights into coupling between widely separate
scales.

The three-body bound state problem we study has a rich history and many
interesting current applications. The continuum limit of this problem
is solved with high precision and the groundwork is laid for the next step
of modulating residual cutoff dependence so that this model can be
applied further.


\begin{acknowledgments}

We are very grateful to Eric Braaten for extensive feedback and many
suggestions. This work was supported in part by the National Science
Foundation  under grants No. PHY-0098645 and No. PHY-0354916. 

\end{acknowledgments}


\end{document}